\documentclass[11pt]{article}
\usepackage{amsmath}
\usepackage[authoryear]{natbib}
\usepackage{latexsym, epsfig, amssymb, amsmath, amsthm, graphicx, mathrsfs}
\usepackage[dvips]{color}
\usepackage[OT1]{fontenc}
\usepackage{hypernat}
\usepackage{anysize}
\renewcommand{\baselinestretch}{1.2}
\marginsize{1.2in}{1in}{0.55in}{1.55in} %
\makeatletter
\def\singlespace{\def\baselinestretch{1}\@normalsize}

\newtheorem{assumption}{Condition}%[section]
\newtheorem{lemma}{Lemma}%[section]
\newtheorem{proposition}{Proposition}%[section]
\newtheorem{theorem}{Theorem}%[section]
\newtheorem{definition}{Definition}%[section]
\makeatother

\newcommand{\ba}{\mbox{\bf a}}
\newcommand{\bb}{\mbox{\bf b}}

\newcommand{\bw}{\mbox{\bf w}}
\newcommand{\bx}{\mbox{\bf x}}
\newcommand{\by}{\mbox{\bf y}}

\newcommand{\bH}{\mbox{\bf H}}

\newcommand{\bX}{\mbox{\bf X}}

\newcommand{\bY}{\mbox{\bf Y}}

\newcommand{\bone}{\mbox{\bf 1}}
\newcommand{\bzero}{\mbox{\bf 0}}
\newcommand{\bveps}{\mbox{\boldmath $\varepsilon$}}

\newcommand{\bbeta}{\mbox{\boldmath $\beta$}}
\newcommand{\bdelta}{\mbox{\boldmath $\delta$}}
\newcommand{\btheta}{\mbox{\boldmath $\theta$}}

\newcommand{\bmu}{\mbox{\boldmath $\mu$}}

\newcommand{\bSig}{\mbox{\boldmath $\Sigma$}}

\newcommand{\hbbeta}{\widehat\bbeta}
\newcommand{\tbbeta}{\widetilde\bbeta}
\newcommand{\tbeta}{\widetilde\beta}

\newcommand{\hbeta}{\widehat{\beta}}

\newcommand{\sbdelta}{\mbox{\scriptsize \boldmath $\delta$}}

\newcommand{\Btau}{{{\cal B}_\tau}}

\newcommand{\Sig}{\mathbf{\Sigma}}
\newcommand{\veps}{\varepsilon}

\newcommand{\diag}{\mathrm{diag}}

\newcommand{\sgn}{\mathrm{sgn}}
\newcommand{\supp}{\mathrm{supp}}

\newcommand{\FS}{\text{FS}}

\def\t{^T}

\begin{document}

\title{Asymptotic Equivalence of Regularization Methods in Thresholded Parameter Space%
\date{}
\author{Yingying Fan and Jinchi Lv %
\thanks{Yingying Fan is Assistant Professor, Information and Operations Management Department, Marshall School of Business, University of Southern California, Los Angeles, CA 90089, USA (e-mail: fanyingy@marshall.usc.edu). Jinchi Lv is Assistant Professor, Information and Operations Management Department, Marshall School of Business, University of Southern California, Los Angeles, CA 90089, USA (e-mail: jinchilv@marshall.usc.edu). Fan's research was supported by NSF CAREER Award DMS-1150318 and Grant DMS-0906784. Lv's research was supported by NSF CAREER Award DMS-0955316 and Grant DMS-0806030. The authors sincerely thank the Co-Editor, Associate Editor, and two referees for their valuable comments that helped improve the paper substantially.
}
\medskip\\
University of Southern California\\
} %
}

\maketitle

\begin{abstract}
High-dimensional data analysis has motivated a spectrum of regularization methods for variable selection and sparse modeling, with two popular classes of convex ones and concave ones. A long debate has been on whether one class dominates the other, an important question both in theory and to practitioners. In this paper, we characterize the asymptotic equivalence of regularization methods, with general penalty functions, in a thresholded parameter space under the generalized linear model setting, where the dimensionality can grow up to exponentially with the sample size. To assess their performance, we establish the oracle inequalities, as in Bickel, Ritov and Tsybakov (2009), of the global minimizer for these methods under various prediction and variable selection losses. These results reveal an interesting phase transition phenomenon. For polynomially growing dimensionality, the $L_1$-regularization method of Lasso and concave methods are asymptotically equivalent, having the same convergence rates in the oracle inequalities. For exponentially growing dimensionality, concave methods are asymptotically equivalent but have faster convergence rates than the Lasso. We also establish a stronger property of the oracle risk inequalities of the regularization methods, as well as the sampling properties of computable solutions. Our new theoretical results are illustrated and justified by simulation and real data examples.
\end{abstract}

\textit{Running title}: Asymptotic Equivalence of Regularization Methods


\section{Introduction} \label{Sec1}
Among all efforts on high-dimensional inference in the last decade, regularization methods have received much attention due to their ability to simultaneously conduct variable selection and estimation. The idea of regularization is to add a penalty term on model complexity to some model fitting loss measure. Then minimizing the penalized model fitting loss measure yields an estimate of the model parameters. Various penalty functions have been proposed in the literature. Broadly speaking, they can be classified into two classes: convex ones and concave ones. The former class is most popularly represented by the Lasso with the $L_1$-penalty \citep{tibshirani2}, and the latter class includes the smoothly clipped absolute deviation (SCAD) \citep{FL01}, minimax concave penalty (MCP) \citep{Zhang07}, and smooth integration of counting and absolute deviation (SICA) \citep{LF09}, among others.

There has been a long debate on which class of regularization methods one should use. Convex regularization methods enjoy nice computational properties and can be efficiently implemented with algorithms such as the LARS \citep{EHJT04} and coordinate optimization \citep{FHHT07,WuLange08}. On the theoretical side, \cite{zhao2} introduced the irrepresentable conditions to characterize the model selection consistency of Lasso. See also, for example, \cite{DET06}, \cite{BTW07}, \cite{vandeGeer08}, and \cite{BRT09} for the properties of the $L_1$-regularization method of Lasso. Despite its appealing properties, the Lasso suffers from an intrinsic bias issue (Fan and Li, 2001; Zou, 2006; Zhang and Huang, 2008). The irrepresentable conditions ensuring the model selection consistency of Lasso become stringent in high dimensions due to increased collinearity among predictors (Lv and Fan, 2009; Fan and Lv, 2011).

On the other hand, concave regularization methods, initiated in \cite{FL01}, ameliorate the bias issue of Lasso and enjoy the model selection consistency property under much weaker conditions. \cite{FL01} proposed nonconcave penalized likelihood methods including the use of the SCAD penalty and established their oracle properties in the finite-dimensional setting. Their results were later extended by \cite{FP04} to the moderate-dimensional setting with $p = o(n^{1/5})$ or $o(n^{1/3})$, where $p$ is the dimensionality and $n$ is the sample size. Recently, \cite{LF09} established the weak oracle properties for regularization methods with general concave penalties in linear regression model, where $p$ is allowed to grow exponentially with sample size $n$. \cite{FL11} extended these results to generalized linear models and further proved the oracle properties of nonconcave penalized likelihood estimators. Despite all these theoretical developments, most existing studies on nonconvex regularization methods
have focused on some appealing local minimizers. The global properties of these methods are still largely unknown and the theoretical characterizations of the global minimizers pose challenges.

The aforementioned advantages and potential issues of the two classes of regularization methods make it difficult for practitioners to decide which one to use. Understanding the connections and differences between different regularization methods is important both theoretically and empirically.  An important question that has long puzzled researchers is: What are the connections and differences of all regularization methods? We intend to provide some answer to this question in this paper. To characterize the performance of different regularization methods, we establish the oracle inequalities and a stronger property of oracle risk inequalities of the global minimizer for regularization methods with general penalty functions, including both convex and concave ones.

The oracle inequalities have been frequently exploited to provide theoretical insights into high-dimensional inference methods and show how closely a sparse modeling method can mimic the oracle procedure. For example, \cite{CandesTao07} proved the oracle inequalities for the Danztig selector, showing that the resulting estimator can achieve a loss within a logarithmic factor of the dimensionality for the oracle estimator. In a seminal paper, \cite{BRT09} established the oracle inequalities simultaneously for two well-known $L_1$-regularization methods, the Lasso and Danztig selector. These oracle inequalities show that the two methods are asymptotically equivalent under certain regularity conditions. Extensive results on the oracle inequalities for general regularization methods were obtained in \cite{Antoniadis01} for the wavelets setting.

Our theoretical analysis reveals the asymptotic equivalence of regularization methods in a thresholded parameter space, in the sense of having the same convergence rates in the oracle inequalities and oracle risk inequalities. The introduction of the thresholded parameter space is motivated by the goal of distinguishing between important predictors and noise predictors in variable selection. The new results on oracle inequalities are parallel to those in Bickel, Ritov and Tsybakov (2009) for the Lasso, but with improved sparsity bound. Our results on the oracle risk inequalities are stronger theoretical developments than those on the oracle inequalities. Specifically, in the case of polynomially growing dimensionality $p$, all regularization methods under consideration including the Lasso and concave ones have the same convergence rates, within a factor of $\log n$ of the oracle rates, in the oracle inequalities and oracle risk inequalities, leading to their asymptotic equivalence. In the case of exponentially growing dimensionality $p$, all concave regularization methods under consideration have the same convergence rates as in the previous case for both oracle inequalities and oracle risk inequalities, but the rates are faster than those of the Lasso, which are within a factor of $\log p$ of the oracle rates.

The connections and differences between the two classes of regularization methods revealed by our study provide an interesting phase transition of how different regularization methods perform as the dimensionality grows with the sample size. To the best of our knowledge, the results and phase transition phenomenon shown in this paper are new to the literature. In addition, our theoretical results are for the global minimizers of the regularization methods, which is different from most studies in the literature.

The rest of the paper is organized as follows. Section \ref{Sec2} introduces the regularization methods in the thresholded parameter space. We present the sampling properties of the concave regularization methods in a thresholded parameter space in ultra-high dimensional generalized linear models, as well as the sampling properties of computable solutions, in Section \ref{Sec3}. We discuss the implementation of the methods and present several simulation and real data examples in Section \ref{Sec5}. Section \ref{Sec6} provides some discussions of our results and their implications. All technical details are relegated to the Appendix.

\section{Regularization methods in thresholded parameter space} \label{Sec2}
Let $(\bx_i, y_i)_{i = 1}^n$ be a sample of $n$ independent observations from $(\bx, Y)$ in the generalized linear model (GLM) linking a $p$-dimensional predictor vector $\bx$ to a scalar response variable $Y$. The GLM assumes that with a canonical link, the conditional distribution of $Y$ given the predictor vector $\bx$ belongs to the exponential family, with a density function taking the form
\begin{align}\label{e016}
f(y; \theta, \phi) = \exp\{y \theta - b(\theta) + c(y, \phi)\},
\end{align}
where $\theta = \bx\t \bbeta$ with $\bbeta = (\beta_1, \cdots, \beta_p)\t \in \mathbb{R}^p$ a regression coefficient vector,  $b(\cdot)$ and $c(\cdot,\cdot)$ are some suitably chosen known functions, and $\phi$ is some positive dispersion parameter. The function $b(\cdot)$ is assumed to be smooth and convex and gives rise to the link function $g(\mu) = \theta$ with $\mu = E(Y|\bx)=b'(\theta)$. Thus the log-likelihood function given by the sample is
\begin{equation} \label{eq:lhood}
\ell_n(\bbeta) = \sum_{i=1}^n \left\{y_i\bx_i^T\bbeta - b(\bx_i^T\bbeta) + c(y_i, \phi)\right\}.
\end{equation}
To ensure model identifiability and improve model interpretability in high dimensions, it is common to assume that only a portion of all predictors contribute to the response, that is, the true regression coefficient vector $\bbeta_0 = (\beta_{0, 1}, \cdots, \beta_{0,p})\t$ is sparse with many components being zero. We refer to predictors with nonzero coefficients $\beta_{0, j}$ as true covariates and the remaining ones as noise covariates. Without loss of generality, we write $\bbeta_0 = (\bbeta_1\t, \bzero\t)\t$ with $\bbeta_1$ consisting of all $s$ nonzero coefficients. To ease the presentation, we suppress the dependence of all parameters such as $s$ and $p$ on $n$ whenever there is no confusion.

In the GLM setting, the regularization method minimizes the penalized negative log-likelihood function
\begin{equation} \label{eq:reg}
Q_n(\bbeta) = -n^{-1}\left\{\by\t\bX\bbeta - \bone\t\bb(\bX\bbeta)\right\} + \|p_\lambda(\bbeta)\|_1,
\end{equation}
where $\by = (y_1, \cdots, y_n)\t$ is an $n$-dimensional response vector, $\bX = (\bx_1, \cdots, \bx_n)\t$ is an $n \times p$ deterministic design matrix, $\bb(\btheta) = (b(\theta_1), \cdots, b(\theta_n))\t$ is a vector-valued function with $\btheta = (\theta_1,\cdots, \theta_n)\t$ and $\theta_i = \bx_i\t\bbeta$, and $\|p_\lambda(\bbeta)\|_1 = \sum_{j=1}^p p_\lambda(|\beta_j|)$ is a separable penalty term on model parameters with $p_\lambda(t)$ a penalty function defined on $t \in [0, \infty)$ and indexed by a nonnegative regularization parameter $\lambda$.   The last term in the log-likelihood function (\ref{eq:lhood}) involving the dispersion parameter $\phi$ is dropped for simplicity. Here we use a compact notation $p_\lambda(\bbeta) = p_\lambda(|\bbeta|) = (p_\lambda(|\beta_1|), \cdots, p_\lambda(|\beta_p|))\t$ with the penalty function applied componentwise and $|\bbeta| = (|\beta_1|, \cdots, |\beta_p|)\t$. To align all covariates to a common scale, we rescale each column vector of the $n \times p$ design matrix $\bX$ for each covariate to have $L_2$-norm $n^{1/2}$. As mentioned in the Introduction, many penalty functions have been proposed for variable selection and sparse modeling; see the references therein for their specific forms.

The level of collinearity among the covariates typically increases with the dimensionality. When this level is high, the estimation can become unstable and the model identifiability may not be guaranteed. We consider the idea of bounding the sparse model size to control the collinearity for sparse models and ensure identifiability and stability of model for reliable prediction and variable selection. A natural bound is given by the following concept of robust spark on the design matrix $\bX$, as introduced in Zheng, Fan and Lv (2012).

\begin{definition}[Robust spark] \label{Def1}
The robust spark $\kappa_c$ of the $n \times p$ design matrix $\bX$ is defined as the smallest possible positive integer such that there exists an $n \times \kappa_c$ submatrix of $n^{-1/2}\bX$ having a singular value less than a given positive constant $c$.
\end{definition}

The above concept of robust spark generalizes that of spark in Donoho and Elad (2003), which plays an important role in the problem of sparse recovery; see also Lv and Fan (2009). As $c\rightarrow 0+$, the robust spark $\kappa_c$ approaches the spark of $\bX$. For each sparse model with size $m < \kappa_c$, the corresponding $n \times m$ submatrix of $n^{-1/2}\bX$ have all singular values bounded from below by $c$. The robust spark $\kappa_c$ is always a positive integer no larger than $n + 1$ and can be some large number diverging with $n$. Although we consider the case of deterministic design matrix, the following proposition formally characterizes the order of $\kappa_c$ when the design matrix $\bX$ is generated from Gaussian distribution.

\begin{proposition}\label{Prop2}
Assume $\log p = o(n)$ and that the rows of the $n \times p$ random design matrix $\bX$ are independent and identically distributed (i.i.d.) as $N(\bzero, \bSig)$, where $\bSig$ has smallest eigenvalue bounded from below by some positive constant. Then there exist positive constants $c$ and $\tilde c$ such that with asymptotic probability one, $\kappa_c \geq \tilde{c} n/(\log p)$.
\end{proposition}

To compare different regularization methods in (\ref{eq:reg}), we introduce the thresholded parameter space
\begin{equation} \label{eq:Bset}
\mathcal{B}_{\tau, c} = \left\{\bbeta \in \mathbb{R}^p: \|\bbeta\|_0 < \kappa_c/2 \text{ and for each }j, \ \beta_j = 0 \text{ or } |\beta_j| \geq \tau\right\},
\end{equation}
where $\bbeta = (\beta_1, \cdots, \beta_p)\t$ and $\tau$ is some positive threshold on parameter magnitude. The threshold $\tau$ is key to distinguishing between important covariates and noise covariates for the purpose of variable selection. As shown in Theorem \ref{Thm1} in Section \ref{Sec3.2}, the threshold $\tau$ is needed to satisfy $\tau \sqrt{n/(\log p)} \rightarrow \infty$ as $n\rightarrow \infty$, indicating that the threshold level should dominate the maximum noise level of $p$ independent standard Gaussian errors asymptotically.

The use of the thresholded parameter space $\mathcal{B}_{\tau, c}$ in (\ref{eq:Bset}) is motivated by the approach of the best subset regression with the $L_0$-regularization, which was proved in Barron, Birge and Massart (1999) to enjoy the oracle risk inequalities under the prediction loss. The following proposition is satisfied by any global minimizer of the regularization problem (\ref{eq:reg}) when the $L_0$-penalty $p_\lambda(t) = \lambda 1_{\{t \neq 0\}}$ is used.

\begin{proposition}[Hard-thresholding property] \label{Prop1}
For the $L_0$-penalty $p_\lambda(t) = \lambda 1_{\{t \neq 0\}}$, the global minimizer $\hbbeta = (\hbeta_1,\cdots, \hbeta_p)\t$ of the regularization problem (\ref{eq:reg}) over $\mathbb{R}^p$ satisfies that each component $\hbeta_j$ is either 0 or has magnitude larger than some positive threshold.
\end{proposition}

The above hard-thresholding property is shared by many other penalty functions. For example, Zheng, Fan and Lv (2012) and Fan and Lv (2012) proved such a property in the setting of penalized least squares for the hard-thresdholding penalty (Hard) and SICA penalty, respectively. These continuous concave penalties are also considered in our study. Intuitively, if some covariates have weak effects, that is, having regression coefficients with magnitude below certain threshold, we can keep these variables out of the model to improve the prediction accuracy with reduced estimation variability because they may have negligible effects on prediction. Moreover, these weak signals are generally difficult to stand out compared with some noise variables due to the impact of high dimensionality.

\section{Asymptotic equivalence of regularization methods} \label{Sec3}
In this section, we establish the asymptotic equivalence of the regularization methods (\ref{eq:reg}) in the thresolded parameter space $\mathcal{B}_{\tau,c}$, with various penalty functions, in the sense of having the same convergence rates in the oracle inequalities and oracle risk inequalities.

\subsection{Technical conditions} \label{Sec3.1}
We first introduce some notation and two key events to facilitate our technical presentation. Denote by $\bveps = (\veps_1,\cdots,\veps_n)\t = \bY - E \bY$ the $n$-dimensional random model error vector with $\bY$ the $n$-dimensional random response vector, and $\alpha_0 = \supp(\bbeta_0) = \{1,\cdots, s\}$ the support of the true regression coefficient vector $\bbeta_0$, that is, the true underlying sparse model. Throughout the paper, we consider a universal choice of the regularization parameter $\lambda = c_0 \sqrt{(\log p)/n}$ with some positive constant $c_0$, where $p$ is implicitly understood as $n \vee p$ in all bounds. Define two events
\begin{equation} \label{eq:Eset}
\mathcal{E} = \left\{\|n^{-1} \bX\t \bveps\|_\infty \leq \lambda/2\right\} \quad \text{ and } \quad \mathcal{E}_0=\left\{\|n^{-1}\bX_{\alpha_0}\t\bveps\|_\infty \leq c_0\sqrt{(\log n)/n}\right\},
\end{equation}
where $\bX_\alpha$ denotes a submatrix of the design matrix $\bX$ consisting of columns with indices in a given set $\alpha \subset \{1, \cdots, p\}$.

\begin{assumption}[Error tail distribution] \label{cond1}
The complements of the two events in (\ref{eq:Eset}) satisfy $P(\mathcal{E}^c) = O(p^{-c_1})$ and $P(\mathcal{E}_0^c) = O(n^{-c_1})$ for some positive constant $c_1$ that can be sufficiently large for large enough $c_0$.
\end{assumption}

\begin{assumption}[Bounded variance] \label{cond2}
The function $b(\theta)$ satisfies that $c_2\leq b''(\theta) \leq c_2^{-1}$ in its domain, where $c_2$ is some positive constant.
\end{assumption}

\begin{assumption}[Concave penalty function] \label{cond3} The penalty function $p_\lambda(t)$ is increasing and concave in $t\in[0,\infty)$ with $p_\lambda(0) = 0$, and is differentiable with $p_{\lambda}'(0+) = c_3 \lambda$ for some positive constant $c_3$.
\end{assumption}

\begin{assumption}[Ultra-high dimensionality]\label{cond4}
It holds that $\log p = O(n^a)$ for some constant $a \in (0,1)$.
\end{assumption}

\begin{assumption}[True parameter vector]\label{cond5}
It holds that $s = o(n^{1-a})$ and there exists a constant $c>0$ such that the robust spark $\kappa_c > 2s$. Moreover, $\min_{1\leq j\leq s}|\beta_{0,j}| \gg \sqrt{(\log p)/n}$.
\end{assumption}

Condition \ref{cond1} puts a constraint on the error tail distribution. The same event $\mathcal{E}$ was considered in Bickel, Ritov and Tsybakov (2009) for Gaussian error, and the probability bound on $P(\mathcal{E}^c)$ can be easily derived using the classical Gaussian tail probability bound. We introduce a second event $\mathcal{E}_0$ to derive improved estimation and prediction bounds for the regularized estimator. The probability bound on $P(\mathcal{E}_0^c)$ holds similarly for Gaussian error.  Condition \ref{cond1} also holds for error distributions other than Gaussian, including bounded or light-tailed error, with no or mild condition on design matrix $\bX$. We discuss some technical details of this condition in Appendix \ref{SecA}.

Condition \ref{cond2} is a mild condition that is commonly assumed in the GLM setting, and requires that the variances of all responses are bounded away from zero and infinity. Condition \ref{cond3} is a common, mild assumption on the penalty function for studying regularization methods; see also Lv and Fan (2009) and Fan and Lv (2011). It requires that the penalty function $p_{\lambda}(t)$ is concave on the positive half axis $[0,\infty)$. In this context, a wide class of penalty functions, including the $L_1$-penalty in Lasso, SCAD, MCP, and SICA, satisfy Condition \ref{cond3} and belong to the class of concave penalty functions.

Condition \ref{cond4} allows the dimensionality $p$ to increase up to exponentially fast with the sample size $n$.  Condition \ref{cond5} puts constraints on the design matrix $\bX$, the model sparsity, and the minimum signal strength. If $\tau$ is chosen such that $\tau \sqrt{n/(\log p)} \rightarrow \infty$ and $\tau <\min_{1\leq j\leq s}|\beta_{0,j}| $, and Condition \ref{cond5} is satisfied, then it is seen that $\bbeta_0 \in \mathcal{B}_{\tau,c}$ with $\mathcal{B}_{\tau,c}$ defined in (\ref{eq:Bset}).   For the reason presented above, in the future presentation, we only consider appropriately chosen $\tau$ such that $\bbeta_0 \in \mathcal{B}_{\tau,c}$. In addition, since we only need the existence of a constant $c$ satisfying Condition \ref{cond5} and its exact value is not needed in implementation,  we will suppress the dependence of $\mathcal{B}_{\tau, c}$ on $c$ and write it as $\mathcal{B}_\tau$ hereafter.

\subsection{Oracle inequalities of global minimizer} \label{Sec3.2}

In this section, we aim to establish the oracle inequalities for the global minimizer of the penalized negative log-likelihood (\ref{eq:reg}) in the thresholded parameter space $\mathcal{B}_{\tau}$, that is,
\begin{equation} \label{e177}
\hbbeta = \arg\min_{\bbeta \in \mathcal{B}_{\tau}}Q_n(\bbeta).
\end{equation}
In general, there may exist multiple global minimizers of $Q_n(\bbeta)$. Our theoretical results are satisfied by any of these global minimizers. Throughout the paper, we refer to any global minimizer as the regularized estimator. The oracle inequalities for the Lasso estimator under estimation and prediction losses were established in Bickel, Ritov and Tsybakov (2009) to study the asymptotic equivalence of the Lasso estimator and Dantzig selector. In addition to common estimation and prediction losses, we introduce a variable selection loss defined as the total number of falsely discovered signs of covariates by an estimator $\hbbeta = (\hbeta_1,\cdots,\hbeta_p)\t$,
\begin{equation} \label{e178}
\FS(\hbbeta) = \left|\left\{j: \sgn(\hbeta_j) \neq \sgn(\beta_{0, j}), 1\leq j \leq p\right\}\right|.
\end{equation}
This loss of false signs $\FS(\hbbeta)$ is a stronger measure than commonly used ones such as the number of false positives and the number of false negatives. We will use this measure to study the sign consistency property of the regularized estimator $\hbbeta$ (Zhao and Yu, 2006).

\begin{theorem}[Oracle inequalities] \label{Thm1}
Assume that Conditions \ref{cond1}--\ref{cond5} hold and $\tau$ is chosen such that $\tau <\min_{1\leq j\leq s}|\beta_{0,j}|$ and $\lambda = c_0 \sqrt{(\log p)/n} = o(\tau)$. Then the global minimizer defined in (\ref{e177}) exists, and any such global minimizer satisfies that with probability at least $1-O(p^{-c_1})$, it holds simultaneously that:
\begin{itemize}
\item[\emph{(a)}] \emph{(False signs)}. $\emph{FS}(\hbbeta) \leq C s \lambda^2 \tau^{-2}/(1 - C \lambda^2 \tau^{-2})$;

\item[\emph{(b)}] \emph{(Estimation losses)}. $\|\hbbeta - \bbeta_0\|_q \leq C \lambda s^{1/q} (1- C \lambda^2 \tau^{-2})^{-1/q}$ for each $q \in [1, 2]$ and $\|\hbbeta - \bbeta_0\|_\infty \leq C \lambda s^{1/2} (1 - C \lambda^2 \tau^{-2})^{-1/2}$;

\item[\emph{(c)}] \emph{(Prediction loss)}. $n^{-1/2}\|\bX(\hbbeta-\bbeta_0)\|_2 \leq C \lambda s^{1/2} (1 - C \lambda^2 \tau^{-2})^{-1/2}$,
\end{itemize}
where $C$ is some positive constant.
\end{theorem}

Theorem \ref{Thm1} shows the existence of the global minimizer defined in (\ref{e177}) and presents the oracle inequalities for the regularized estimator for a wide class of penalty functions characterized by Condition \ref{cond3}. All theoretical results in the paper hold uniformly over the set of all possible global minimizers.

Since the regularization parameter $\lambda$ represents the minimum regularization level needed to suppress the noise covariates, and the thresholding level $\tau$ is just below the minimum signal strength, a valid thresholding level requires $\lambda = o(\tau)$ to ensure that all true covariates will not be screened out asymptotically. Since $\lambda \tau^{-1} \rightarrow 0$, the above bound on false signs $\FS(\hbbeta)$ is of a smaller order than the true model size $s$, meaning that the proportion of missed signs for signals, that is, $\FS(\hbbeta)/s$, vanishes asymptotically. This tight bound on false signs is a unique feature of introducing the thresholded parameter space. In contrast, the bound on estimated model size $\|\hbbeta\|_0$ for the ordinary Lasso estimator is of order $O(\phi_{\max} s)$ with $\phi_{\max}$ the largest eigenvalue of the Gram matrix $n^{-1} \bX\t \bX$ (Bickel, Ritov and Tsybakov, 2009), and thus the proportion of missed signs $\FS(\hbbeta)/s$ in this estimator can be of order $O(\phi_{\max})$ which does not vanish asymptotically. In view of $\lambda \tau^{-1} \rightarrow 0$ and  $\lambda = c_0\sqrt{(\log p)/n}$, the bounds on the estimation and prediction losses in Theorem \ref{Thm1} satisfy that for each $q\in [1,2]$,
\begin{align*}
\|\hbbeta - \bbeta_0\|_q = O\left\{s^{1/q}\sqrt{(\log p)/n}\right\} \quad \text{and} \quad n^{-1/2}\|\bX(\hbbeta-\bbeta_0)\|_2 = O(\sqrt{s(\log p)/n}),
\end{align*}
whose convergence rates are within a logarithmic factor of $\log p$ of the oracle rates. The above convergence rates in these oracle inequalities are consistent with those in Bickel, Ritov and Tsybakov (2009) for the Lasso estimator.

We next show that under some additional conditions, the sign consistency of the regularized estimator $\hbbeta$ can be obtained and the convergence rates in Theorem \ref{Thm1} can be further improved. Define a small neighborhood of $\bbeta_0$ in the thresholded parameter space as
\begin{equation} \label{e047}
\mathcal{B}_1^* = \left\{\bbeta \in \mathcal{B}_{\tau}: \supp(\bbeta)=\supp(\bbeta_0) \text{ and } \|\bbeta - \bbeta_0\|_2 \leq 2 C s^{1/2} \lambda\right\}
\end{equation}
with constant $C$ given in Theorem \ref{Thm1}. Note that this neighborhood is asymptotically shrinking since $s^{1/2} \lambda \rightarrow 0$ as guaranteed by Conditions \ref{cond4} and \ref{cond5}. We introduce two important constants
\begin{align} \label{e059}
&\gamma_n^* = \sup_{\bbeta_i\in \mathcal{B}_1^*, \ i=1,\cdots, n} \left\|\Big\{\frac{1}{n}\bX_{\alpha_0}\t\bH(\bbeta_1,\cdots, \bbeta_n)\bX_{\alpha_0}\Big\}^{-1}\right\|_\infty,  \\
\label{e059b}
& \gamma_n = \sup_{\bbeta\in \mathcal{B}_1^*,\ \alpha\subset\{s+1,\cdots,p\}\text{ and } |\alpha|\leq s }\left\|\frac{1}{n}\bX_{\alpha_0}\t\bH(\bbeta)\bX_{\alpha}\right\|_\infty,
\end{align}
where  $\bH(\bbeta_1,\cdots, \bbeta_n) = \diag\{b''(\bx_1\t\bbeta_1), \cdots, b''(\bx_n\t\bbeta_n)\}$ and $\bH(\bbeta) = \diag\{b''(\bx_1\t\bbeta), \cdots, b''(\bx_n\t\bbeta)\}$ are diagonal matrices of variances. To get some intuition on the constants $\gamma_n^*$ and $\gamma_n$, let us consider the special case of Gaussian linear model with $b''(\theta) \equiv 1$. In such case we have
 \begin{equation} \label{e065}
\gamma_n^* = \left\|\Big(\frac{1}{n}\bX_{\alpha_0}\t\bX_{\alpha_0}\Big)^{-1}\right\|_\infty \quad \text{ and } \quad \gamma_n = \sup_{ \alpha\subset\{s+1,\cdots,p\}\text{ and } |\alpha|\leq s}\left\|\frac{1}{n}\bX_{\alpha_0}\t\bX_{\alpha}\right\|_\infty.
\end{equation}
Since each column of $\bX$ is rescaled to have $L_2$-norm $n^{1/2}$, it is seen that $\gamma_n^*$ is only associated with the design matrix of the true model $\alpha_0$, while $\gamma_n$ is related to the correlation between true covariates and noise covariates.

To evaluate the prediction property, we consider the Kullback-Leibler (KL) divergence of the fitted model from the true model given by
\[
D(\hbbeta) = -(E \bY)\t\bX(\hbbeta - \bbeta_0) + \bone\t \left[\bb(\bX\hbbeta) - \bb(\bX\bbeta_0)\right],
\]
where $E \bY = (b'(\bx_1^T\bbeta_0), \cdots, b'(\bx_n^T\bbeta_0))\t$ is the true mean response vector for the GLM.

\begin{theorem}[Sign consistency and oracle inequalities] \label{Thm2}
Assume that conditions of Theorem \ref{Thm1} hold with $\min_{1\leq j\leq s}|\beta_{0, j}| \geq 2\tau$, $\lambda =c_0\sqrt{(\log p)/n}= o (s^{-1/2} \tau)$, and $\gamma_n = o \big\{\tau\sqrt{n/( s\log n)}\big\}$. Then any global minimizer $\hbbeta$ in (\ref{e177}) satisfies that with probability at least $1-O(n^{-c_1})$, it holds simultaneously that:
\begin{itemize}
\item[\emph{(a)}] \emph{(Sign consistency)}. $\sgn(\hbbeta) = \sgn(\bbeta_0)$;

\item[\emph{(b)}] \emph{(Estimation and prediction losses)}. If the penalty function further satisfies $p_\lambda'(\tau) = O\big\{\sqrt{(\log n)/n}\big\}$, then we have for each $q \in [1,2]$,
    \[ \|\hbbeta - \bbeta_{0}\|_{q}\leq Cs^{1/q}\sqrt{(\log n)/n}, \quad \|\hbbeta - \bbeta_{0}\|_{\infty}\leq C\gamma_{n}^*\sqrt{(\log n)/n}, \]
    and $n^{-1}D(\hbbeta) \leq C s (\log n)/n$,
\end{itemize}
where $C$ is some positive constant.
\end{theorem}

In comparison with Theorem \ref{Thm1}(a), we obtain in Theorem \ref{Thm2}(a) a stronger property of sign consistency of the regularized estimator. The additional condition on the penalty function $p_\lambda'(\tau) = O\big\{\sqrt{(\log n)/n}\big\}$ can be easily satisfied by concave penalties such as the SCAD and SICA, with appropriately chosen $\lambda$. For penalty functions satisfying this additional condition, the convergence rates of the regularized estimator are improved with the $\log p$ term (see Theorem \ref{Thm1}) replaced with $\log n$ (see Theorem \ref{Thm2}). In this sense, our study provides a setting showing the general nonoptimality of the logarithmic factor of the dimensionality $\log p$ in oracle inequalities.

To gain more insights into Theorem \ref{Thm2}, we consider again the case of Gaussian linear model. In view of (\ref{e065}) and the robust spark condition in (\ref{eq:Bset}), we have an upper bound on $\gamma_n^*$ given by
\[
\gamma_n^* \leq s^{1/2}  \left\|\Big(\frac{1}{n}\bX_{\alpha_0}\t\bX_{\alpha_0}\Big)^{-1}\right\|_2 \leq c^{-1} s^{1/2}. \]
Observing that $\gamma_n$ in (\ref{e065}) measures the correlation between noise covariates and true covariates, the condition $\gamma_n = o \big\{\tau\sqrt{n/( s\log n)}\big\}$ in Theorem \ref{Thm2} essentially requires that the noise covariates and true covariates should not be too highly correlated with each other. Note that each column of $\bX_{\alpha_0}$ is rescaled to have $L_2$-norm $n^{1/2}$. When all true covariates are orthogonal to each other, we have $\gamma_n^* = 1$ and thus the bound on the $L_\infty$-estimation loss in Theorem \ref{Thm2} becomes
\[
\|\hbbeta - \bbeta_{0}\|_{\infty}\leq C \sqrt{(\log n)/n},
\]
whose convergence rate is within a logarithmic factor of $\log n$ of the oracle rate.

Combining Theorems \ref{Thm1} and \ref{Thm2} shows that for polynomially growing dimensionality with $p = O(n^a)$ for some positive constant $a$, the $L_1$-regularization method of Lasso and concave regularization methods with penalties satisfying Condition \ref{cond3} are asymptotically equivalent in the thresholded parameter space, meaning that all methods have the same convergence rates in the oracle inequalities, with a logarithmic factor of $\log n$. For exponentially growing dimensionality with $\log p = O(n^a)$ for some positive constant $a$ less than $1$, the concave regularization methods satisfying the additional condition $p_\lambda'(\tau) = O\big\{\sqrt{(\log n)/n}\big\}$ are asymptotically equivalent and still enjoy the same convergence rates in the oracle inequalities, with a logarithmic factor of $\log n$. For the $L_1$-penalty used in Lasso, the condition $p_\lambda'(\tau)  = O\big\{\sqrt{(\log n)/n}\big\}$ and the choice of the regularization parameter $\lambda = c_0 \sqrt{(\log p)/n}$ are, however, incompatible with each other in the case of $\log p = O(n^a)$. Thus in the ultra-high dimensional case, the convergence rates in the oracle inequalities for Lasso, which have a logarithmic factor of $\log p$, are slower than those for concave regularization methods. These results reveal an interesting phase diagram on how the performance of regularization methods, in the thresholded parameter space, evolves with the dimensionality and the penalty function, in terms of convergence rates in the oracle inequalities.

Among different approaches to alleviating the bias issue of the Lasso, the adaptive Lasso \citep{Zou06} exploits the weighted $L_1$-penalty $\lambda \|\bw \circ \bbeta\|_1$ with weight vector $\bw = (w_1, \cdots, w_p)\t$, where $w_j = |\beta_{\text{ini}, j}|^{-\gamma}$ for some $\gamma > 0$, $1 \leq j \leq p$, with $\bbeta_{\text{ini}} = (\beta_{\text{ini}, 1}, \cdots, \beta_{\text{ini}, p})\t$ an initial estimator, and $\circ$ denotes the componentwise product. Under some particular choices of the initial estimator, the adaptive Lasso can enjoy the properties established in Theorems \ref{Thm1} and \ref{Thm2}, similarly as the Lasso. For instance, the choice of the trivial initial estimator $\bbeta_{\text{ini}} = \bone$ gives the Lasso estimator. How to choose other nontrivial initial estimators is crucial to ensuring that the adaptive Lasso has improved convergence rates as concave methods in ultra high dimensions. Another popular method, the bridge regression in \cite{FF93}, uses the $L_q$-penalty $p_\lambda(t) = \lambda t^q$ for $0 < q \leq 2$. When $0 < q < 1$, the bridge regression is also a concave regularization method since $p_\lambda(t)$ is concave on $[0, \infty)$. However, such a method falls outside the class of regularization methods in our framework, since $p_{\lambda}'(0+) = \infty$ in this case which violates Condition \ref{cond3}. As a consequence, a key inequality (\ref{e025}) in our technical analysis does not hold in general for the bridge estimator with $0 < q < 1$. It is yet unclear whether similar results to those in Theorems \ref{Thm1} and \ref{Thm2} would also hold for the bridge estimator in the case of $0 < q < 1$.

\subsection{Oracle risk inequalities of global minimizer} \label{Sec3.3}
The oracle inequalities presented in Section \ref{Sec3.2} are derived by conditioning on the event $\mathcal{E}$ (Theorem \ref{Thm1}) or $\mathcal{E}\cap \mathcal{E}_0$ (Theorem \ref{Thm2}) defined in (\ref{eq:Eset}), and thus they may not hold on the complement $\mathcal{E}^c$ or $\mathcal{E}^c\cup \mathcal{E}_0^c$. We now establish a stronger property of the oracle risk inequalities for the regularized estimator $\hbbeta$ in (\ref{e177}), which gives upper bounds on the expectations of various variable selection, estimation, and prediction losses.

\begin{theorem}[Oracle risk inequalities] \label{Thm3}
Assume that conditions of Theorem \ref{Thm2} hold and the fourth moments of errors $E \veps_i^4$ are uniformly bounded. Then any global minimizer $\hbbeta$ in (\ref{e177}) satisfies that:
\begin{itemize}
\item[\emph{(a)}] \emph{(Sign risk)}. $E \big\{\emph{FS}(\hbbeta)\big\} = \frac{1}{p_{\lambda }(\tau)}\big\{ [\|p_{\lambda }(\bbeta_0)\|_1 + s\lambda^2] O(n^{-c_1}) + O(p^{-c_1/2}\kappa_c)\big\}$;

\item[\emph{(b)}] \emph{(Estimation and prediction risks)}. If the penalty function further satisfies $p_\lambda'(\tau) = O\big\{\sqrt{(\log n)/n}\big\}$, then we have for each $q \in [1,2]$,
        \[ E \|\hbbeta-\bbeta_0\|_q^q \leq C s \big[(\log n)/n\big]^{q/2}, \quad E \|\hbbeta-\bbeta_0\|_\infty \leq C\gamma_n^*\sqrt{(\log n)/n}, \]
    and $E \big\{n^{-1}D(\hbbeta)\big\} \leq C s (\log n)/n$,
\end{itemize}
where $C$ is some positive constant.
\end{theorem}

The expectation of the number of falsely discovered signs converges to zero at a polynomial rate of $n$. In the wavelet setting of Gaussian linear model with $p=n$ and orthogonal design matrix $\bX$, it has been proved in \cite{Antoniadis01} that the risks of the regularized estimators under the $L_2$-loss are bounded by $O\{s(\log n)/n\}$, which is consistent with our results above.  This indicates that there is no additional cost in risk bounds for generalizing to the ultra-high dimensional nonlinear model setting of GLM.

\subsection{Sampling properties of computable solutions}
The theoretical results presented in previous sections are on any global minimizer of the penalized negative log-likelihood $Q_n(\bbeta)$ in the thresholded parameter space $\mathcal{B}_{\tau}$. The global minimizer may not be guaranteed to be found by a computational algorithm. Therefore, it is also important to study the sampling properties of the computable solution produced by any algorithm. Define a vector-valued function $\bmu(\btheta) = (b'(\theta_1), \cdots, b'(\theta_n))\t$ for $\btheta = (\theta_1, \cdots, \theta_n)\t$, which is the mean function in the GLM.

\begin{theorem} \label{Thm4}
Let $\hbbeta \in \mathcal{B}_{\tau}$ be a computable solution to the minimization problem (\ref{e177}) produced by any algorithm that is the global minimizer when constrained on the subspace given by $\supp(\hbbeta)$, and $\eta_n = \|n^{-1}\bX\t[\by - \bmu(\bX\hbbeta)]\|_\infty$. Assume in addition that there exists some positive constant $c_4$ such that $\|n^{-1} \bX_\alpha\t [\bmu(\bX \bbeta) - \bmu(\bX \bbeta_0)]\|_2 \geq c_4 \|\bbeta - \bbeta_0\|_2$ for any $\bbeta \in \mathcal{B}_{\tau}$ and $\alpha = \supp(\bbeta) \cup \supp(\bbeta_0)$, if the model (\ref{e016}) is nonlinear. If $\eta_n + \lambda=o(\tau)$ and $\min_{1\leq j\leq s}|\beta_{0,j}| > c_5 s^{1/2} (\eta_n+\lambda)$ with $c_5$ some sufficiently large positive constant, then $\hbbeta$ enjoys the same asymptotic properties as for any global minimizer in Theorems \ref{Thm1}--\ref{Thm3} under the same conditions therein.
\end{theorem}

The condition that $\hbbeta$ is the global minimizer of the problem (\ref{e177}) when constrained on the subspace given by its support can hold under some mild condition on the penalty function. Such a property has been formally characterized in Proposition 1 of Fan and Lv (2011). For example, when condition (\ref{e076}) in Section \ref{Sec4} is satisfied, the penalized negative log-likelihood $Q_n(\bbeta)$ in (\ref{eq:reg}) is strictly convex on the above subspace, which entails that the local minimizer found by any algorithm will necessarily be the global minimizer over this subspace.

As shown in the proof of Theorem \ref{Thm4}, the above additional condition on the mean deviation vector $\bmu(\bX \bbeta) - \bmu(\bX \bbeta_0)$ always holds for linear model with $c_4 = c^2$. In nonlinear models, such a condition requires that a deviation from the true mean vector $\bmu(\bX \bbeta_0)$ can be captured by the covariates involved. Theorem \ref{Thm4} shows that a computable solution produced by any algorithm can share the same nice asymptotic properties as for any global minimizer, when the maximum correlation between the covariates and the residual vector $\by - \bmu(\bX\hbbeta)$ is a smaller order of the threshold $\tau$. Such a solution needs not to be the global minimizer.

\section{Numerical studies} \label{Sec5}
\subsection{Implementation} \label{Sec4}
Algorithms for implementing regularization methods include those mentioned in the Introduction, the LQA algorithm (Fan and Li, 2001), and LLA algorithm (Zou and Li, 2008). In particular, the coordinate optimization algorithm, which solves the problem one coordinate a time and cycles through all coordinates, has received much recent attention for solving large-scale problems thanks to its very low computational cost for each coordinate. For example, the ICA algorithm (Fan and Lv, 2011) implements regularization methods by combining the ideas of second-order quadratic approximation of likelihood function and coordinate optimization. For each coordinate within each iteration, the quadratic approximation of the likelihood function at the $p$-vector from the previous step along that coordinate reduces the problem to a univariate penalized least squares, which admits a closed-form solution for many commonly used penalty functions. See, for example, Lin and Lv (2013) for an analysis of convergence properties of this algorithm.

In this paper, we apply the ICA algorithm to implement concave regularization methods in the thresholded parameter space. A key ingredient of these methods is the use of the thresholded parameter space, which naturally puts an additional constraint on each component of the parameter vector. For each coordinate within each iteration, we solve the univariate penalized least-squares problem with the corresponding quadratic approximation of the likelihood function, and update this coordinate only when the global minimizer has magnitude above the given threshold $\tau$. We found that this optimization algorithm works well for producing the solution paths for concave regularization methods in the thresholded parameter space. The thresholding also induces additional sparsity of the regularized estimate and thus makes the algorithm converge faster.

To gain some insight into the stability of the computational algorithm, assume that the penalty function $p_\lambda(t)$ has maximum concavity
\begin{equation} \label{e076}
\rho(p_\lambda) = \sup_{0 < t_1 < t_2 < \infty} \left\{- \frac{p_\lambda'(t_2) - p_\lambda'(t_1)}{t_2-t_1}\right\} < c c_2,
\end{equation}
where constants $c$ and $c_2$ are given in Definition \ref{Def1} and Condition \ref{cond2}, respectively. This condition holds for penalties satisfying Condition \ref{cond3} with suitably chosen regularization parameter $\lambda$ and shape parameter. For example, the $L_1$-penalty $p_{\lambda}(t) = \lambda t$ in Lasso has maximum concavity $0$, the SCAD penalty $p_\lambda(t)$ having derivative $p_\lambda'(t)=\lambda I(t\le\lambda)+ (a-1)^{-1} (a\lambda-t)_+ I(t>\lambda)$, with shape parameter $a > 2$, has maximum concavity $\rho(p_\lambda) = (a-1)^{-1}$, and the SICA penalty $p_\lambda(t; a) = \lambda (a + 1)t/(a + t)$ with shape parameter $a$ has maximum concavity $2 \lambda (a^{-1} + a^{-2})$. Condition (\ref{e076}) on the maximum concavity of penalty function ensures that the penalized negative log-likelihood $Q_n(\bbeta)$ in (\ref{eq:reg}) is strictly convex on a union of coordinate subspaces $\{\bbeta \in \mathbb{R}^p: \|\bbeta\|_0 < \kappa_c\}$, which is key to the stability of the sparse solution found by any algorithm.

In implementation, we need to select two tuning parameters: the threshold $\tau$ for the thresholded parameter space $\mathcal{B}_\tau$ and the regularization parameter $\lambda$ for the penalty function $p_\lambda(t)$. As shown in the theoretical results, the threshold $\tau$ should be larger than the regularization parameter $\lambda  = c_0 \sqrt{(\log p)/n}$ in order to filter the noise. Thus we choose $\tau$ as $\tau = c_6 (\log n)^{1/2} \sqrt{ (\log p)/n}$ for some positive constant $c_6$. As for the regularization parameter $\lambda$, we use the validation set or cross-validation to select $\tau$.

\subsection{Simulation studies} 
In this section, we investigate the finite-sample properties of several concave regularization methods in the thresholded parameter space, in three commonly used generalized linear models: the linear regression model, the logistic regression model, and the Poisson regression model, as well as in a real data example. Since the main purpose of our simulation study is to justify the theoretical results, we select the tuning parameters by minimizing the prediction error calculated using an independent validation set, with size equal to the sample size in the study. This tuning parameter selection criterion reduces additional estimation variability incurred by the cross-validation (CV). Fivefold CV was used for tuning parameter selection in real data analysis.

\subsubsection{Linear regression} \label{Sec5.1.1}
We start with the linear regression model (\ref{e016}) written in the matrix form
\begin{equation} \label{e044}
\by = \bX \bbeta + \bveps.
\end{equation}
We generated 100 data sets from this model with error $\bveps \sim N(\bzero, \sigma^2 I_n)$ independent of the design matrix $\bX$. The sample size $n$ and error standard deviation $\sigma$ were chosen to be $100$ and $0.4$, respectively. For each data set, the rows of the design matrix $\bX$ were sampled as i.i.d. copies of random $p$-vector from $N(\bzero, \Sig)$ with $\Sig = (r^{|j - k|})_{1 \leq j, k \leq p}$ for some number $r$. We considered three settings for the pair $(p, r)$ of dimensionality and population collinearity level: $(1000, 0.25)$, $(1000, 0.5)$, and $(5000, 0.25)$. In addition to the population collinearity, the sample collinearity among the covariates can be of a much higher level due to the high dimensionality. The true regression coefficient vector $\bbeta$ was set to be $\bbeta_0 = (1, -0.5, 0.7, -1.2, -0.9, 0.5, 0.55, 0, \cdots, 0)\t$. We take the oracle procedure, using the information of the true underlying sparse model, as the benchmark variable selection method, and compare the Lasso, SCAD, MCP, Hard, and SICA in the thresholded parameter space, which are referred to as Lasso$_t$, SCAD$_t$, MCP$_t$, Hard$_t$, and SICA$_t$ for simplicity, respectively. We also include the original SCAD in comparison. Simulation results show that SCAD$_t$, MCP$_t$, and Hard$_t$ had very similar performance, so we omit the results on MCP$_t$ and Hard$_t$ to save space. The shape parameter $a$ of the SCAD and SICA penalties was chosen to be $3.7$, and $10^{-4}$ or $10^{-2}$, respectively.

\begin{table}[!htb]
{\small
\caption{The means and standard errors (in parentheses) of various performance measures as well as the estimated error standard deviation for all methods in Section \ref{Sec5.1.1}; settings I, II, and III refer to cases of $(p, r) = (1000, 0.25)$, $(1000, 0.5)$, and $(5000, 0.25)$, respectively}
\centering
\begin{tabular}{lccccc}
\hline
Measure & \multicolumn{4}{c}{Method} \\
\cline{2-6}
 & Lasso$_t$ & SCAD & SCAD$_t$ & SICA$_t$ & Oracle \\
\cline{2-6}
Setting I &  &   &  &  & \\
PE ($\times 0.1$) & 1.722 (0.007) & 1.736 (0.007)  & 1.721 (0.007) & 1.719 (0.007) & 1.719 (0.007) \\
$L_2$-loss ($\times 0.1$) & 1.122 (0.032) & 1.184 (0.030) & 1.115 (0.030)  &  1.106 (0.031) & 1.106 (0.031) \\
$L_1$-loss ($\times 0.1$) & 2.485 (0.077) & 2.972 (0.100) & 2.425 (0.071)  & 2.414 (0.071) & 2.414 (0.071) \\
$L_\infty$-loss ($\times 0.01$) & 7.48 (0.24) & 7.67 (0.21)  & 7.61 (0.21)  & 7.55 (0.23) & 7.55 (0.23) \\
FP & 0.01 (0.01) & 3.84 (0.47) &  0 (0)  & 0 (0) & 0 (0) \\
FN & 0 (0) & 0 (0) &  0 (0)  & 0 (0) & 0 (0) \\
$\widehat{\sigma}$ ($\times 0.1$) & 4.040 (0.035) & 3.959 (0.034)  & 4.019 (0.034) & 4.011 (0.034) & 4.011 (0.034) \\
\cline{2-6}
Setting II &  &  &    &  & \\
PE ($\times 0.1$) & 1.789 (0.045) & 1.741 (0.008)  & 1.735 (0.008)  & 1.738 (0.019) & 1.719 (0.007) \\
$L_2$-loss ($\times 0.1$) & 1.445 (0.100) &  1.403 (0.039) & 1.375 (0.040) &  1.353 (0.062) & 1.301 (0.038) \\
$L_1$-loss ($\times 0.1$) & 3.360 (0.318) & 3.558 (0.118)  & 3.180 (0.108)  & 2.957(0.132) & 2.862 (0.088) \\
$L_\infty$-loss ($\times 0.01$) & 9.42 (0.65) & 8.99 (0.28)  & 8.95 (0.28)  &  9.22 (0.49) & 8.76 (0.26) \\
FP & 0.22 (0.18) &  4.11 (0.48) & 0.56  (0.12)  & 0.01 (0.01) & 0 (0) \\
FN & 0.01 (0.01) & 0 (0)  & 0 (0)  & 0.01 (0.01) & 0 (0) \\
$\widehat{\sigma}$ ($\times 0.1$) & 4.023 (0.033) & 3.937 (0.034)  & 3.963 (0.036)  & 4.016 (0.035) & 4.010 (0.034) \\
\cline{2-6}
Setting III &  &  &    &  & \\
PE ($\times 0.1$) & 1.722 (0.008) & 1.743 (0.008)  & 1.719 (0.007)  & 1.724 (0.008) & 1.715 (0.006) \\
$L_2$-loss ($\times 0.1$) & 1.133 (0.034) &  1.228 (0.033) & 1.123 (0.032)  & 1.138 (0.034) & 1.104 (0.031) \\
$L_1$-loss ($\times 0.1$) & 2.457 (0.074) & 3.455 (0.139)  & 2.455 (0.071)  & 2.488 (0.075) &  2.438 (0.070) \\
$L_\infty$-loss ($\times 0.01$) & 7.77 (0.28) & 7.79 (0.24)  & 7.61 (0.26)  & 7.80 (0.29) & 7.43 (0.24) \\
FP & 0.02 (0.01) & 8.25 (0.84)  & 0.01 (0.01)  & 0.06 (0.02) & 0 (0) \\
FN & 0 (0) & 0 (0)  & 0 (0)  & 0 (0) & 0 (0) \\
$\widehat{\sigma}$ ($\times 0.1$) & 4.003 (0.032) & 3.859 (0.034)  & 3.988 (0.031)  & 3.966 (0.032) & 3.983 (0.031) \\
\hline
\end{tabular}
\label{tab1}
}
\end{table}

\begin{table}[!htb]
\small
\caption{Model selection consistency probabilities of all methods in Section \ref{Sec5.1.1}}
\centering
\begin{tabular}{lccccc}
\hline
Setting of $(p, r)$ & \multicolumn{5}{c}{Model selection consistency probability} \\
\cline{2-6}
 & Lasso$_t$ & SCAD & SCAD$_t$ & SICA$_t$ & Oracle \\
\cline{2-6}
$(1000, 0.25)$ & 0.99 & 0.26 & 1  & 1  & 1 \\
$(1000, 0.5)$ & 0.96 & 0.26 & 0.71  & 0.98 & 1 \\
$(5000, 0.25)$ & 0.98 & 0.14 & 0.99  & 0.94 & 1 \\
\hline
\end{tabular}
\label{tab2}
\end{table}

\begin{table}[!htb]
{\small
\caption{The means and standard errors (in parentheses) of various performance measures as well as the estimated error standard deviation for all methods in Section \ref{Sec5.1.1} with $(p, r) = (5000, 0.5)$; settings I, II, and III refer to cases of $n = 100$, $200$, and $400$, respectively}
\centering
\begin{tabular}{lccccc}
\hline
Measure & \multicolumn{4}{c}{Method} \\
\cline{2-6}
 & Lasso$_t$ & SCAD & SCAD$_t$ & SICA$_t$ & Oracle \\
\cline{2-6}
Setting I &  &   &  &  & \\
PE ($\times 0.1$) & 2.584 (0.215) & 1.958 (0.105)  & 1.820 (0.062) & 2.103 (0.134) & 1.715 (0.006) \\
$L_2$-loss ($\times 0.1$) & 2.935 (0.343) & 1.824 (0.187) & 1.555 (0.126)  &  2.102 (0.243) & 1.304 (0.039) \\
$L_1$-loss ($\times 0.1$) & 6.750 (0.841) & 5.296 (0.523) & 3.681 (0.280)  & 4.618 (0.535) & 2.909 (0.089) \\
$L_\infty$-loss ($\times 0.01$) & 19.28 (2.21) & 11.48 (1.25)  & 10.02 (0.91)  & 14.31 (1.68) & 8.63 (0.29) \\
FP & 0.19 (0.07) & 11.33 (1.00) &  0.91 (0.17)  & 0.08 (0.03) & 0 (0) \\
FN & 0.41 (0.09) & 0.06 (0.03) &  0.05 (0.04)  & 0.21 (0.07) & 0 (0) \\
$\widehat{\sigma}$ ($\times 0.1$) & 4.394 (0.111) & 3.893 (0.061) & 3.930 (0.050) & 4.169 (0.082) & 3.983 (0.031) \\
\cline{2-6}
Setting II &  &  &    &  & \\
PE ($\times 0.1$) & 1.655 (0.004) & 1.661 (0.004) & 1.654 (0.004)  & 1.652 (0.004) & 1.652 (0.004) \\
$L_2$-loss ($\times 0.1$) & 0.920 (0.034) & 0.951 (0.032)  & 0.916 (0.031) &  0.891 (0.031) & 0.894 (0.031) \\
$L_1$-loss ($\times 0.1$) & 2.025 (0.079) &  2.427 (0.141) & 1.996 (0.071)  & 1.952(0.071) & 1.958 (0.072) \\
$L_\infty$-loss ($\times 0.01$) & 6.08 (0.22) &  6.21 (0.22) & 6.19 (0.22)  &  5.98(0.22) & 6.00 (0.22) \\
FP & 0 (0) & 4.82 (1.23)  & 0  (0)  & 0 (0) & 0 (0) \\
FN & 0 (0) &  0 (0) & 0 (0)  & 0 (0) & 0 (0) \\
$\widehat{\sigma}$ ($\times 0.1$) & 4.021 (0.020) & 3.970 (0.023)  & 4.012 (0.020)  & 4.010 (0.020) & 4.010 (0.020) \\
\cline{2-6}
Setting III &  &  &    &  & \\
PE ($\times 0.1$) & 1.626 (0.003) & 1.629 (0.003)  & 1.626 (0.003)  & 1.625 (0.003) & 1.625 (0.003) \\
$L_2$-loss ($\times 0.1$) & 0.676 (0.020) &  0.692 (0.020) & 0.673 (0.019)  & 0.661 (0.019) & 0.665 (0.019) \\
$L_1$-loss ($\times 0.1$) & 1.505 (0.048) & 1.713 (0.084)  & 1.489 (0.045)  & 1.469 (0.044) &  1.473 (0.044) \\
$L_\infty$-loss ($\times 0.01$) & 4.39 (0.13) &  4.43 (0.13) & 4.42 (0.13)  & 4.31 (0.13) & 4.36 (0.13) \\
FP & 0 (0) & 3.67 (1.11)  & 0 (0)  & 0 (0) & 0 (0) \\
FN & 0 (0) & 0 (0)  & 0 (0)  & 0 (0) & 0 (0) \\
$\widehat{\sigma}$ ($\times 0.1$) & 4.009 (0.012) & 3.993 (0.013)  & 4.008 (0.012)  & 4.007 (0.012) & 4.006 (0.012) \\
\hline
\end{tabular}
\label{tab9}
}
\end{table}

\begin{table}[!htb]
\small
\caption{Model selection consistency probabilities of all methods in Section \ref{Sec5.1.1} with $(p, r) = (5000, 0.5)$}
\centering
\begin{tabular}{cccccc}
\hline
$n$ & \multicolumn{5}{c}{Model selection consistency probability} \\
\cline{2-6}
 & Lasso$_t$ & SCAD & SCAD$_t$ & SICA$_t$ & Oracle \\
\cline{2-6}
100 & 0.78 & 0.10 &  0.68  & 0.84  & 1 \\
200 & 1 & 0.55 & 1  & 1 & 1 \\
400 & 1 & 0.69 & 1  & 1 & 1 \\
\hline
\end{tabular}
\label{tab10}
\end{table}

\begin{figure}[!htb] \centering
\begin{center}%
\begin{tabular}
[l]{l}%
{\hspace{-0.4in}\includegraphics[scale=0.75]%
{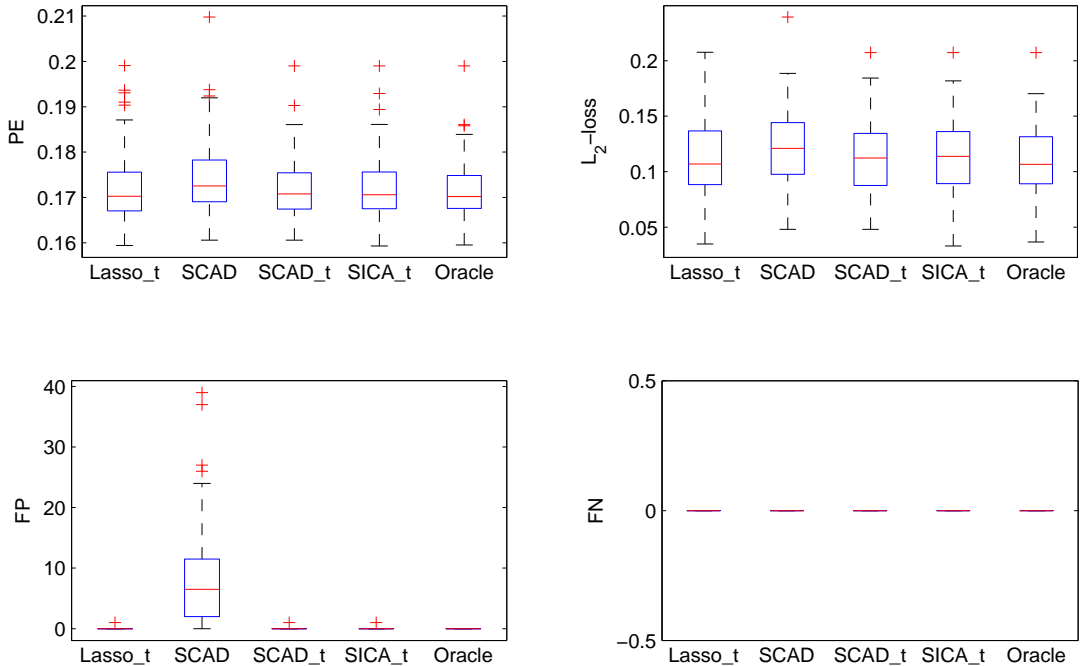}%
}%
\end{tabular}%
\vspace{-0.25in}
\caption{Boxplots of the PE, $L_2$-loss, FP, and FN over $100$ simulations for all methods in Section \ref{Sec5.1.1}, with $(p, r) = (5000, 0.25)$. The $x$-axis represents different methods.}
\label{Fig1}%
\end{center}%
\end{figure}%

To evaluate the selected models, we consider several performance measures for prediction and variable selection. The first measure is the prediction error (PE) defined as $E (Y - \bx\t \hbbeta)^2$ with $\hbbeta$ an estimate and $(\bx\t, Y)$ an independent observation for the $p$ covariates and response. An independent test sample of size $10,000$ was generated to calculate the PE. The second to fourth measures are the $L_q$-estimation losses $\|\hbbeta - \bbeta_0\|_q$ with $q = 2, 1$, and $\infty$, respectively. The fifth and sixth measures are variable selection losses of false positives (FP) and false negatives (FN), where a false positive represents a falsely selected noise covariate in the model and a false negative represents a missed true covariate. The seventh measure is the model selection consistency probability of each method based on 100 simulations. We also compare the estimate $\widehat{\sigma}$ of the error standard deviation $\sigma$ in linear model for all methods.

The model selection consistency results are summarized in Table \ref{tab2} and all other results are summarized in Table \ref{tab1}. We see that across all settings and over all performance measures through their means and standard errors, all concave regularization methods in the thresholded parameter space mimicked very closely the oracle procedure. In particular, the model selection consistency probability for each of these methods was very close to one and its estimated error standard deviation followed very closely that by the oracle procedure. Figure \ref{Fig1} further shows that the regularized estimators given by these methods had almost identical sampling distributions. These numerical results are in line with our theory presented in Section \ref{Sec3}. We also observe that SCAD$_t$ improved over original SCAD in both prediction and variable selection. The model selection consistency probability of SCAD was particularly improved when considering the thresholded parameter space.

We also consider three additional settings for $(n, p, r)$: $(100, 5000, 0.5)$, $(200, 5000, 0.5)$, and $(400, 5000, 0.5)$. The comparison results for all methods are presented in Tables \ref{tab9} and \ref{tab10}. Due to the high collinearity in the setting of $(n, p, r) = (100, 5000, 0.5)$, all methods performed worse than the oracle procedure. As sample size increases, these methods followed more closely the oracle procedure, which are consistent with our theoretical results.

\subsubsection{Logistic regression} \label{Sec5.1.2}
We consider the logistic regression model (\ref{e016}) with the parameter $\theta_i$ for the response $Y_i$ given by
\begin{equation} \label{e045}
\btheta = (\theta_1, \cdots, \theta_n)\t = \bX \bbeta.
\end{equation}
We generated 100 data sets from this model, each of which contains an $n$-dimensional response vector $\by$ sampled from the Bernoulli distribution with success probability vector $(e^{\theta_1}/(1+e^{\theta_1}), \cdots, e^{\theta_n}/(1+e^{\theta_n}))\t$, where $\btheta$ is given in (\ref{e045}). The sample size $n$ and  the true regression coefficient vector $\bbeta$ were set to be $200$ and $\bbeta_0 = (2, 0, -2.3, 0, 2.8, 0, -2.2, 0, 2.5, 0, \cdots, \\0)\t$, respectively. The rest of the setting is the same as that in Section \ref{Sec5.1.1}. We compared the same concave regularization methods with the oracle procedure and used the same seven prediction and variable selection performance measures as in Section \ref{Sec5.1.1}. The prediction error is defined as $E \{Y - \exp(\bx\t \hbbeta)/[1+\exp(\bx\t \hbbeta)]\}^2$ with $\hbbeta$ an estimate and $(\bx\t, Y)$ an independent observation for the $p$ covariates and response.

\begin{table}[!htb]
{\small \caption{The means and standard errors (in parentheses) of various prediction and variable selection performance measures for all methods in Section \ref{Sec5.1.2}; settings I, II, and III refer to cases of $(p, r) = (1000, 0.25)$, $(1000, 0.5)$, and $(5000, 0.25)$, respectively}
\centering
\begin{tabular}{lccccc}
\hline
Measure & \multicolumn{5}{c}{Method} \\
\cline{2-6}
& Lasso$_t$ & SCAD & SCAD$_t$ & SICA$_t$ & Oracle \\
\cline{2-6}
Setting I &  &   &  &  & \\
PE ($\times 0.01$) & 7.89 (0.03) & 7.97 (0.07)  & 7.86 (0.03) & 7.88 (0.04) & 7.86 (0.03) \\
$L_2$-loss & 0.954 (0.039) & 1.033 (0.096)  &  0.915 (0.052)  &  0.913 (0.051) & 0.897 (0.049) \\
$L_1$-loss & 1.927 (0.087) &  2.130 (0.271) & 1.793 (0.108)  & 1.788 (0.107) & 1.757 (0.103) \\
$L_\infty$-loss ($\times 0.1$) & 6.354 (0.224) & 6.936 (0.509)  & 6.346 (0.345)  & 6.348 (0.346) & 6.238 (0.333) \\
FP & 0.02 (0.01) &  0.09 (0.05) & 0 (0) & 0.01 (0.01) & 0 (0) \\
FN & 0 (0) &  0 (0) & 0 (0)  & 0 (0) & 0 (0) \\
\cline{2-6}
Setting II &  &  &  &  & \\
PE ($\times 0.01$) & 9.09 (0.07) & 9.13 (0.07)  & 9.00 (0.05)  &  9.04 (0.06) & 8.94 (0.03) \\
$L_2$-loss & 1.002 ( 0.059) &  0.998 (0.072)   & 0.916 (0.059) &  0.908 (0.055) & 0.855 (0.049) \\
$L_1$-loss & 2.044 (0.135) &   2.036 (0.168)  &  1.824 (0.129) &  1.802 ( 0.120) & 1.678 (0.103) \\
$L_\infty$-loss ($\times 0.1$) & 6.549 (0.334) & 6.574 (0.408)   & 6.213 (0.360) &  6.154 (0.338) & 5.926 (0.314) \\
FP & 0.06 (0.02) &  0.16 (0.05)  & 0.04 (0.02)  & 0.06 (0.03) & 0 (0) \\
FN & 0.01 (0.01) & 0 (0)  & 0 (0)  & 0 (0) & 0 (0) \\
\cline{2-6}
Setting III &  &   &  &  & \\
PE ($\times 0.01$) & 7.89 (0.04) & 7.96 (0.07)   & 7.94 (0.07) & 7.98 (0.08) & 7.85 (0.04) \\
$L_2$-loss & 1.060 (0.053) &  1.200 (0.089) &  1.172 (0.083)  &  1.175 (0.084) & 1.102 (0.079) \\
$L_1$-loss & 2.156 (0.123) & 2.411 (0.199) &   2.337 (0.181)  & 2.335 (0.182) &   2.200 (0.172) \\
$L_\infty$-loss ($\times 0.1$) & 6.935 (0.301) & 8.157 (0.568)  & 7.997 (0.547)  & 8.091 (0.560) & 7.495 (0.501) \\
FP & 0.03 (0.02) & 0.03 (0.02)  & 0.01 (0.01) & 0.01 (0.01) & 0 (0) \\
FN & 0 (0) & 0.02 (0.01)  & 0.02 (0.01) & 0.03 (0.02) & 0 (0) \\
\hline
\end{tabular}
\label{tab3}
}
\end{table}

\begin{table}[!htb]
\small
\caption{Model selection consistency probabilities of all methods in Section \ref{Sec5.1.2}}
\centering
\begin{tabular}{lccccc}
\hline
Setting of $(p, r)$ & \multicolumn{5}{c}{Model selection consistency probability} \\
\cline{2-6}
 & Lasso$_t$ & SCAD & SCAD$_t$ & SICA$_t$ & Oracle \\
\cline{2-6}
$(1000, 0.25)$ & 0.98 & 0.95  & 1   & 0.99 & 1 \\
$(1000, 0.5)$ & 0.93 & 0.89 & 0.96   & 0.95  & 1 \\
$(5000, 0.25)$ & 0.97 & 0.95 & 0.97  & 0.96 & 1 \\
\hline
\end{tabular}
\label{tab4}
\end{table}

\begin{figure}[!htb] \centering
\begin{center}%
\begin{tabular}
[l]{l}%
{\hspace{-0.4in}\includegraphics[scale=0.75]%
{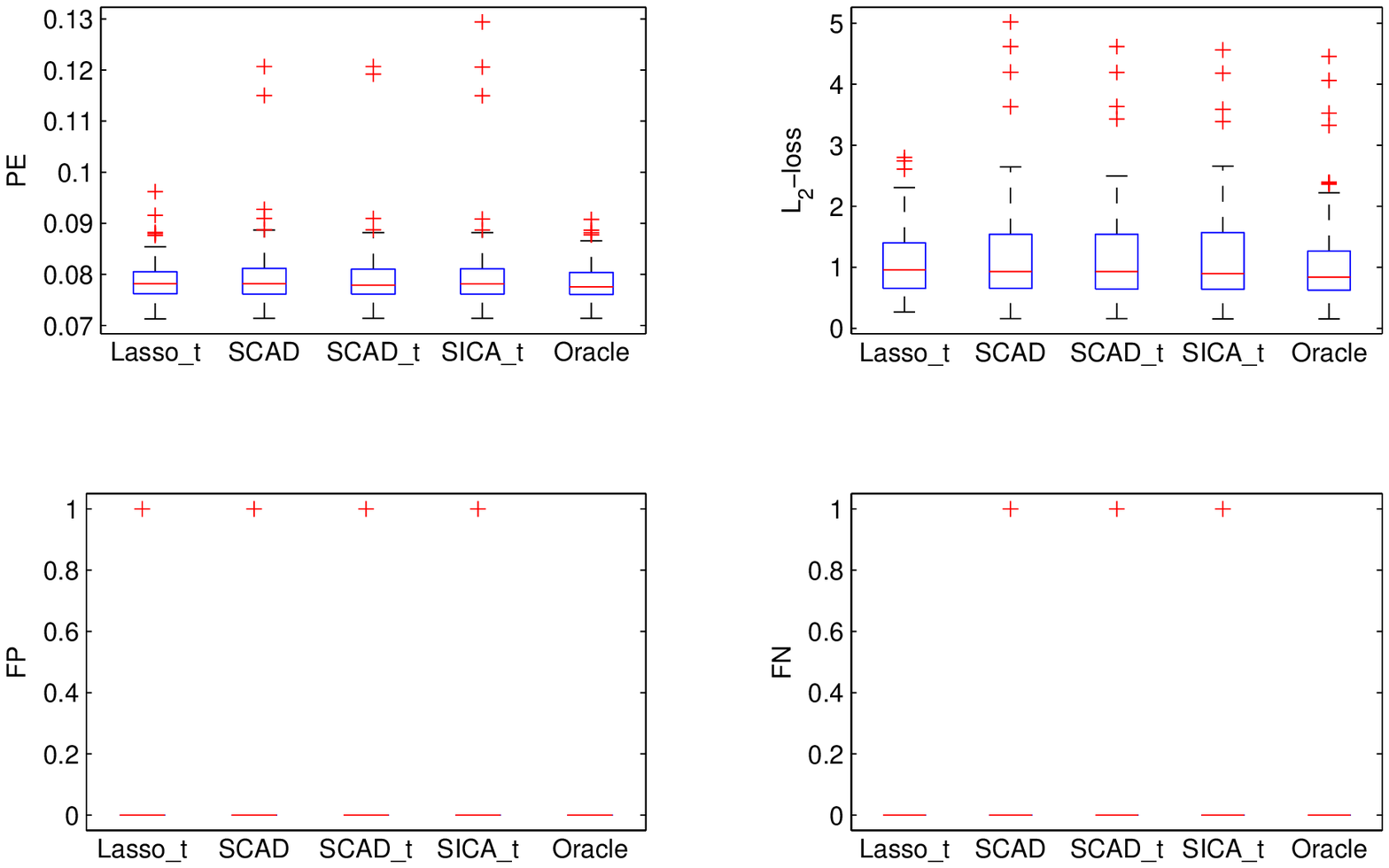}%
}%
\end{tabular}%
\vspace{-0.25in}
\caption{Boxplots of the PE, $L_2$-loss, FP, and FN over $100$ simulations for all methods in Section \ref{Sec5.1.2}, with $p = 5000$. The $x$-axis represents different methods.}
\label{Fig2}%
\end{center}%
\end{figure}%

Tables \ref{tab3}--\ref{tab4} and Figure \ref{Fig2} summarize the comparison results of all methods. The conclusions are similar to those in Section \ref{Sec5.1.1}. Facilitated by the thresholded parameter space, all methods mimicked very closely the oracle procedure in this nonlinear model for binary data, confirming the theoretical results.

\subsubsection{Poisson regression} \label{Sec5.1.3}
We now consider the Poisson regression model (\ref{e016}) with the parameter $\theta_i$ for the response $Y_i$ given as in (\ref{e045}). We generated 100 data sets from this model, each of which contains an $n$-dimensional response vector $\by$ sampled from the Poisson distribution with mean vector $(e^{\theta_1}, \cdots, e^{\theta_n})\t$, where $\btheta = (\theta_1, \cdots, \theta_n)\t$ is given in (\ref{e045}). The sample size $n$ and  the true regression coefficient vector $\bbeta$ were set to be $200$ and $\bbeta_0 = (1, -0.9, 0.8, -1.1, 0.6, 0, \cdots, 0)\t$, respectively. The rest of the setting is the same as that in Section \ref{Sec5.1.2}. We compared the same concave regularization methods with the oracle procedure, using the same seven prediction and variable selection performance measures as in Section \ref{Sec5.1.1}. The prediction error is defined as $E [Y - \exp(\bx\t \hbbeta)]^2$ with $\hbbeta$ an estimate and $(\bx\t, Y)$ an independent observation for the $p$ covariates and response.

\begin{table}[!htb]
{\small \caption{The 5\% trimmed means and standard errors (in parentheses) of various prediction and variable selection performance measures for all methods in Section \ref{Sec5.1.3}; settings I, II, and III refer to cases of $(p, r) = (1000, 0.25)$, $(1000, 0.5)$, and $(5000, 0.25)$, respectively}
\centering
\begin{tabular}{lccccc}
\hline
Measure & \multicolumn{5}{c}{Method} \\
\cline{2-6}
& Lasso$_t$ & SCAD & SCAD$_t$ & SICA$_t$ & Oracle \\
\cline{2-6}
Setting I &  &   &  &  & \\
PE & 21.34 (2.23) &  13.11 (0.94)  & 9.00 (0.66)  & 7.39 (0.50) & 6.22 (0.22) \\
$L_2$-loss ($\times 0.01$) & 19.62 (1.44) & 17.09 (0.71)  & 11.58 (0.67) &  9.05 (0.45) & 7.94 (0.31) \\
$L_1$-loss ($\times 0.1$) & 4.658 (0.440) & 4.714 (0.191)  & 2.513 (0.173)  & 1.720 (0.086) & 1.513 (0.060) \\
$L_\infty$-loss ($\times 0.01$) & 12.13 (0.75) & 11.39 (0.58)  & 7.82 (0.43)  & 6.69 (0.37) & 5.69 (0.23) \\
FP & 1.47 (0.28) &  11.61 (0.67) & 1.20 (0.22)  & 0.07 (0.03) & 0 (0) \\
FN & 0 (0) & 0 (0)  & 0 (0)  & 0 (0) & 0 (0) \\
\cline{2-6}
Setting II &  &  &  &  & \\
PE & 5.934 (0.272) &  3.288 (0.065)  & 2.903 (0.055)  & 2.754 (0.072) & 2.655 (0.030) \\
$L_2$-loss ($\times 0.01$) & 38.34 (1.78) & 19.12 (0.60)   & 14.75 (0.69)  &  12.52 (0.82) & 11.54 (0.48) \\
$L_1$-loss ($\times 0.1$) & 10.07 (0.651) & 5.860 (0.225)  & 3.254 (0.192) & 2.357(0.143) & 2.180 (0.097) \\
$L_\infty$-loss ($\times 0.01$) & 20.96 (0.75) &  12.21 (0.49)  & 9.92 (0.45) &  9.15 (0.69) & 8.39 (0.35) \\
FP & 2.54 (0.23) & 14.84 (0.84)   & 1.70 (0.34) & 0 (0) & 0 (0) \\
FN & 0 (0) &  0 (0)  & 0 (0) & 0.01 (0.01) & 0 (0) \\
\cline{2-6}
Setting III &  &   &  &  & \\
PE & 34.86 (3.04) &  14.26 (1.00)  & 10.38 (0.71) & 8.28 (0.51) & 6.15 (0.20) \\
$L_2$-loss ($\times 0.01$) & 31.67 (1.92) &  18.42 (0.55)  & 14.41 (0.76) & 11.62 (0.72) & 8.35 (0.31) \\
$L_1$-loss ($\times 0.1$) & 8.158 (0.617) &  5.863 (0.175)  & 3.600 (0.260) & 2.252(0.173) &  1.552 (0.062) \\
$L_\infty$-loss ($\times 0.01$) & 18.29 (0.88) &  12.10 (0.48)  & 9.46 (0.46) & 8.48 (0.46) & 6.21 (0.23) \\
FP & 2.36 (0.28) & 19.12 (0.71)   & 3.06 (0.49) & 0.39 (0.08) & 0 (0) \\
FN & 0 (0) &  0 (0)  & 0 (0) & 0 (0) & 0 (0) \\
\hline
\end{tabular}
\label{tab5}}
\end{table}

\begin{table}[!htb]
\caption{Model selection consistency probabilities of all methods in Section \ref{Sec5.1.3}}
\centering
\begin{tabular}{lccccc}
\hline
Setting of $(p, r)$ & \multicolumn{5}{c}{Model selection consistency probability} \\
\cline{2-6}
 & Lasso$_t$ & SCAD & SCAD$_t$ & SICA$_t$ & Oracle \\
\cline{2-6}
$(1000, 0.25)$ & 0.51 & 0  & 0.65  & 0.89 & 1 \\
$(1000, 0.5)$ & 0.17 & 0  & 0.68  & 0.94 & 1 \\
$(5000, 0.25)$ & 0.27 & 0 & 0.51  & 0.72 & 1 \\
\hline
\end{tabular}
\label{tab6}
\end{table}

\begin{figure}[!htb] \centering
\begin{center}%
\begin{tabular}
[l]{l}%
{\hspace{-0.4in}\includegraphics[scale=0.75]%
{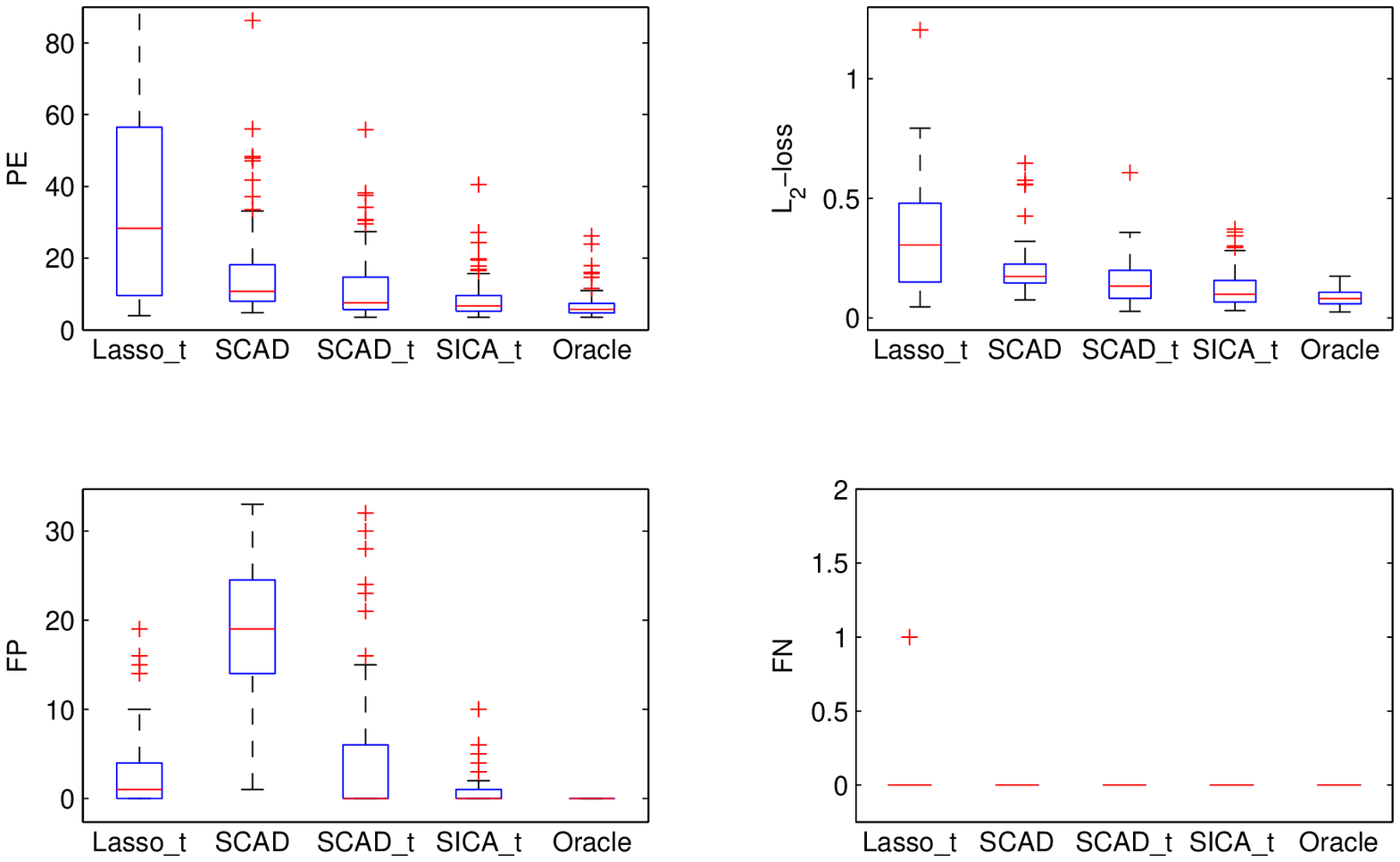}%
}%
\end{tabular}%
\vspace{-0.25in}
\caption{Boxplots of the PE, $L_2$-loss, FP, and FN over $100$ simulations for all methods in Section \ref{Sec5.1.3}, with $p = 5000$. The $x$-axis represents different methods.}
\label{Fig3}%
\end{center}%
\end{figure}%

Tables \ref{tab5}--\ref{tab6} and Figure \ref{Fig3} summarize the comparison results for all methods. As shown in Figure \ref{Fig3}, the boxplot for the prediction error of the oracle procedure exhibits some outliers. This is caused by the random design matrix which may not be well-behaved in some samples, leading to some unstable estimates of the true regression coefficients. The instability comes from the fact that the variance of a Poisson random variable is equal to its mean and thus is generally unbounded if the mean is not bounded. To better compare the performance of all methods in such a case, we considered the 5\% trimmed means, excluding 5\% of values from each tail, and their standard errors of different prediction and variable selection measures. The asymptotic equivalence of concave regularization methods in the thresholded parameter space shown in the theory was also demonstrated in this nonlinear model for count data. But compared to linear models, the finite-sample performance of these methods differs more from that of the oracle procedure, indicating the increased difficulty of model inference for nonlinear models. The improvement of SCAD$_t$ over original SCAD was more profound in this setting.

\begin{table}[!htb]
\caption{The means and standard errors of classification errors by different methods over 50 random splittings of the prostate cancer data in Section \ref{Sec5.2}}
\centering
\begin{tabular}{lcccc}
\hline
   & Lasso$_t$ & SCAD & SCAD$_t$ & SICA$_t$ \\
\hline
    Mean & 1.42  & 4.36 & 3.44  & 1.30  \\
    Standard error & 0.20 & 0.36  & 0.32 & 0.18 \\
\hline
\end{tabular}
\label{tab7}
\end{table}

\begin{table}[!htb]
\caption{Selection probabilities of most frequently selected genes with number up to median model size by each method across 50 random splittings of the prostate cancer data in Section \ref{Sec5.2}}
\centering
\begin{tabular}{lcccclcccc}
\hline
Gene ID & Lasso$_t$ & SCAD & SCAD$_t$ & SICA$_t$ & Gene ID & Lasso$_t$ & SCAD & SCAD$_t$ & SICA$_t$ \\
\hline
1018  & ---  &  ---   & ---     & 0.44  &  7139  & 0.38  &  ---   & ---     & 0.52     \\
4525  & 0.96  &  ---   & ---     & 0.94  &  7539  & 0.94  &  ---   & ---     & 0.94 \\
4636  & 0.42  &  ---   & ---     & 0.40  &  8123  & 0.44  &  0.12   & ---     & 0.42 \\
5319  & 0.54  &  ---   & ---     & 0.64  &  9093  & 0.86  &  0.12   & 0.08     & 0.90 \\
5661  & 0.68  &  ---   & 0.12     & 0.64  & 9126  & ---  &  0.10   & ---     & --- \\
5890  & 1  &  0.10   & 0.28     & 1  & 10292  & 0.36  &  ---   & ---     & 0.40 \\
5977  & 0.58  &  ---   & ---     & 0.44  &  10494  & 0.80  &  ---   & ---     & 0.74 \\
6145  & 0.28  &  ---   & ---     & 0.26  &  10537  & 0.82  &  ---   & ---     & 0.74 \\
6185  & 0.94  &  0.10   & 0.10     & 0.98 &  11215  & 0.32  &  ---   & ---     & 0.32  \\
6390  & 0.28  &  ---   & ---     & ---  &  11871  & 1  &  ---   & 0.24     & 1 \\
6462  & 0.36  &  ---   & ---     & ---  & 12547  & 0.28  &  ---   & ---     & 0.28  \\
6512  & 0.48  &  ---   & ---     & 0.46 \\
\hline
\end{tabular}
\label{tab8}
\end{table}

\subsection{Real data example} \label{Sec5.2}
We apply all the methods to the prostate cancer data set which was originally studied in Singh et al. (2002) and is available at http://www.broadinstitute.org/cgi-bin/cancer/datasets.cgi. This data set, which was also analyzed in Fan and Fan (2008), consists of 136 patient samples with 77 from the prostate tumor group (labeled as $1$) and 59 from the normal group (labeled as $0$). For each patient, we have the gene expression measurements for 12,600 genes.

Following Singh et al. (2002) and Fan and Fan (2008), we randomly split the 136 samples into a training set of 52 samples from the cancer class and 50 samples from the normal class, and a test set of 25 samples from the cancer class and 9 samples from the normal class. For each splitting of the data set, we fit the logistic regression model to the training data with the regularization methods. We then calculated the classification error using the test data. We repeated the random splitting 50 times, and the means and standard errors of classification errors are summarized in Table \ref{tab7}. We also calculated the median model size by each method: 21 by Lasso$_t$, 5 by SCAD, 5 by SCAD$_t$, and 20 by SICA$_t$. For each method, we computed the percentage of times each gene was selected and listed the most frequently chosen $m$ genes in Table \ref{tab8}, with $m$ equal to the median model size by the method. We see that Lasso$_t$ and SICA$_t$ performed similarly, and SCAD and SCAD$_t$ produced more sparse models than the other two methods.

\section{Discussions} \label{Sec6}
We have studied the asymptotic equivalence of two popular classes of regularization methods with convex penalties and concave penalties, in high-dimensional generalized linear models. Our framework covers many commonly used regularization methods such as the Lasso and concave ones such as the SCAD, MCP, and SICA. The oracle inequalities as well as the stronger property of the oracle risk inequalities of the global minimizer for the regularization methods have been established to characterize their connections and differences. When the Lasso penalty is considered, our oracle inequalities are consistent with those in Bickel, Ritov and Tsybakov (2009), with improved sparsity thanks to the introduced thresholded parameter space. The established theoretical results have revealed an interesting phenomenon of phase transition in both linear and nonlinear models, confirmed by our numerical studies. We have also established additional theoretical results to provide insights into the sampling properties of computable solutions.

To simplify the technical presentation and better illustrate the ideas, we have focused on the setting of generalized linear models and the Lasso for the convex class of regularization methods. The theoretical results in the paper may hold in more general model settings as well. The phase transition phenomenon may also be shown for other convex penalties such as the $L_2$-penalty. These problems are beyond the scope of the current paper and will be interesting topics for future research.

\appendix

\section{Technical details on Condition \ref{cond1}} \label{SecA}
We show that the two probability bounds in Condition \ref{cond1} hold for a wide class of error distributions. To this end, note that an application of the Bonferroni inequality gives
\begin{equation} \label{e019}
P\big(\|n^{-1} \bX\t \bveps\|_\infty > \lambda/2\big) \leq \sum\nolimits_{j=1}^p P(n^{-1}|\widetilde{\bx}_j\t\bveps|>\lambda/2),
\end{equation}
where $(\widetilde{\bx}_1, \cdots, \widetilde{\bx}_p) = \bX$. We consider two cases of error distribution.

\textit{Case 1}: Bounded error. Assume that $|\veps_i| \leq a$ for each $1 \leq i \leq n$, with $a$ being some positive constant. Then it follows from Hoeffding's inequality (Hoeffding, 1963) that
\begin{equation} \label{e021}
P(n^{-1} |\widetilde{\bx}_j\t\bveps|>\lambda/2) \leq 2 \exp\left(-\lambda^2 n/(8 a^2)\right),
\end{equation}
since $\|\widetilde{\bx}_j\|_2 = n^{1/2}$ for each $j$. This probability bound is of order $O\big\{p^{-c_0^2/(8 a^2)}\big\}$ when $\lambda = c_0 \sqrt{(\log p)/n}$.

\textit{Case 2}: Light-tailed error. Assume that there exist positive constants $M, v_0$ such that
\begin{equation} \label{e022}
E \left[\exp(M^{-1} |\veps_i|) - 1 - M^{-1} |\veps_i|\right] M^2 \leq v_0/2
\end{equation}
holds uniformly for $1 \leq i \leq n$. Then it follows from Bernstein's inequality (Bennett, 1962; van der Vaart and Wellner, 1996) that
\begin{equation} \label{e023}
P(n^{-1} |\widetilde{\bx}_j\t\bveps|>\lambda/2) \leq 2 \exp\left(-\frac{\lambda^2 n}{8 v_0 + 4 \|\widetilde{\bx}_j\|_\infty M \lambda}\right),
\end{equation}
since $\|\widetilde{\bx}_j\|_2 = n^{1/2}$ for each $j$. This probability bound is of order $O\big\{p^{-c_0^2/(8 v_0 + d)}\big\}$ with $d = 4 c_0 M \|\widetilde{\bx}_j\|_\infty \sqrt{(\log p)/n}$ when $\lambda = c_0\sqrt{(\log p)/n}$, and this bound becomes $O\big\{p^{-c_0^2/(12 v_0)}\big\}$ if we further assume $\|\widetilde{\bx}_j\|_\infty \leq (c_0 M)^{-1} v_0\sqrt{n/(\log p)}$. This additional assumption means that the maximum absolute element of the design matrix $\bX$ is bounded from above by $(c_0 M)^{-1} v_0\sqrt{n/(\log p)}$, which is a mild condition. Condition (\ref{e022}) was also made in Fan and Lv (2011) for analyzing nonconcave penalized likelihood estimators in GLM, and is mild in view of the moment-generating function of distributions in the exponential family.

When $c_0$ is large enough, combining the above two cases with (\ref{e019}) leads to the desired probability bound on $P(\mathcal{E}^c)$ in Condition \ref{cond1}. Since $|\alpha_0| = s \leq n$ is assumed implicitly, similar probability bounds hold for the event $\mathcal{E}_0^c$. Thus we impose Condition \ref{cond1} instead of making explicit assumptions on the model error distribution and design matrix $\bX$.

\section{Proofs of main results} \label{SecB}
For notational simplicity, we use $C$ to denote a generic positive constant, whose value may change from line to line. Denote by $\bb'(\btheta) = (b'(\theta_1), \cdots, b'(\theta_n))\t$ the $n$-vector of mean function, $\bb''(\btheta) = (b''(\theta_1),\cdots,b''(\theta_n))\t$ the $n$-vector of variance function for $\btheta = (\theta_1 ,\cdots, \theta_n)\t \in \mathbb{R}^n$, and $\ba_{\alpha}$ the subvector of a vector $\ba\in \mathbb{R}^p$ formed by components with indices in a given set $\alpha\subset\{1,\cdots, p\}$.

\subsection{Proof of Proposition \ref{Prop2}} \label{SecB.1}
Let $k_0 = c_4 n$ be an integer with $c_4 \in (0,1)$ some constant. For each set $\alpha_1 \subset\{1,\cdots, p\}$ with $|\alpha_1| = k_0$, denote by $\bSig_{\alpha_1,\alpha_1}$ the principal submatrix of $\bSig$ corresponding to variables in $\alpha_1$. We will show that there exist some universal positive constants $c_5$ and $C_1$ such that
\begin{align}\label{e279}
P\left\{\lambda_{\min}\big(n^{-1}\bX_{\alpha_1}\t\bX_{\alpha_1} \big) < c_5\right\} \leq  \exp(-C_1n),
\end{align}
where $\lambda_{\min}(\cdot)$ denotes the smallest eigenvalue of a matrix. Note that for any submatrix $n^{-1/2}\bX_{\alpha}$ with $|\alpha| \leq k_0$, its smallest singular value is bounded from below by the smallest singular value of  $n^{-1/2}\bX_{\alpha_1}$ with $|\alpha_1|= k_0$ and $\alpha_1\supset \alpha$. It follows that $\lambda_{\min}\big(n^{-1}\bX_{\alpha}\t\bX_{\alpha} \big)$ satisfies the same deviation probability bound (\ref{e279}). Thus, an application of the Bonferroni inequality with $K$ an integer satisfying $K = 2^{-1} C_1 n/(\log p) \leq k_0$ gives
\[
P\left\{\min_{|\alpha|\leq K} \lambda_{\min}\big(n^{-1}\bX_{\alpha}\t\bX_{\alpha} \big) < c_5\right\} \leq  \sum_{|\alpha| \leq K}\exp(-C_1n) \leq p^{K}\exp(-C_1n) \rightarrow 0.
\]
This shows that with asymptotic probability one, $\kappa_c \geq K$ for any $c \leq c_5$.

It remains to prove (\ref{e279}). Define $\widetilde\bX_{\alpha_1} =\bX_{\alpha_1}\bSig_{\alpha_1,\alpha_1}^{-1/2}$. Then $\bX_{\alpha_1}\t\bX_{\alpha_1} = \bSig^{1/2}_{\alpha_1,\alpha_1}\widetilde\bX_{\alpha_1}\t\widetilde\bX_{\alpha_1}\bSig^{1/2}_{\alpha_1,\alpha_1} $ and the rows of $\widetilde\bX_{\alpha_1}$ are i.i.d. standard Gaussian random vectors. Since $\bSig$ has smallest eigenvalue bounded from below, we have
\begin{align*}
\lambda_{\min}\big(n^{-1}\bX_{\alpha_1}\t\bX_{\alpha_1}\big) \geq \lambda_{\min}\big(n^{-1}\widetilde\bX_{\alpha_1}\t\widetilde\bX_{\alpha_1}\big)\lambda_{\min}(\bSig_{\alpha_1,\alpha_1})\geq C\lambda_{\min}\big(n^{-1}\widetilde\bX_{\alpha_1}\t\widetilde\bX_{\alpha_1}\big).
\end{align*}
So we only need to show that $\lambda_{\min}\big(n^{-1}\widetilde\bX_{\alpha_1}\t\widetilde\bX_{\alpha_1}\big)$ satisfies a similar deviation probability bound as (\ref{e279}), which is entailed by the concentration property proved in Fan and Lv (2008) (see their deviation inequality (16)). This completes the proof.

\subsection{Proof of Proposition \ref{Prop1}} \label{SecB.2}
Since $\hbbeta = (\hbeta_1,\cdots, \hbeta_p)\t$ is the global minimizer of $Q_n(\bbeta)$, it holds that for each $j$,  $\hbeta_j$ is also the global minimizer of the same objective function along the $j$-th coordinate, that is, $\hbeta_j$ minimizes
\[ \widetilde{Q}_n(\beta_j) = a_0 -n^{-1}\by\t \widetilde{\bx}_j\beta_j + \sum_{i=1}^nb(a_i +  x_{ij}\beta_j) + \lambda 1_{\{|\beta_j|\neq 0\}}, \]
where $(\widetilde{\bx}_1,\cdots, \widetilde{\bx}_p) = \bX$ with $\widetilde{\bx}_j = (x_{1j}, \cdots, x_{nj})\t$ and $a_i$'s with $i=0,1,\cdots, n$ are constants independent of $\beta_j$. Note that the first three terms of $\widetilde{Q}_n(\beta_j)$ are continuous functions of $\beta_j$, while the last term is a step function of $\beta_j$.  Thus it follows easily that the global minimizer $\hbeta_j$ is either 0 or has magnitude larger than certain positive threshold whose value depends on $\lambda $ and the continuous part of $\widetilde{Q}_n(\beta_j)$, which concludes the proof.

\subsection{Lemma \ref{Lem1} and its proof}  \label{SecB.3}
We single out a lemma that is used in the proofs of Theorems \ref{Thm1}--\ref{Thm3}.

\begin{lemma} \label{Lem1}
Under Conditions \ref{cond2}--\ref{cond3}, we have
\begin{equation}\label{e026}
\|\bdelta\|_2^2 \leq c^{-1}n^{-1}\|\bX\bdelta\|_2^2 \leq C \big(\|n^{-1}\bX\t \bveps\|_{\infty} + \lambda \big)\|\bdelta\|_1,
\end{equation}
where $\bdelta = \hbbeta - \bbeta_0$ is the estimation error for the regularized estimator $\hbbeta$ in (\ref{e177}), $c$ is the positive constant in Definition \ref{Def1}, and $C$ is some positive constant.
\end{lemma}

\noindent \textit{Proof of Lemma \ref{Lem1}}. Since $\hbbeta$ is the global minimizer of $Q_n(\bbeta)$ in $\mathcal{B}_\tau$ and $\bbeta_0 \in \mathcal{B}_\tau$, it follows that
\begin{equation} \label{f001}
0\leq Q_n(\bbeta_0) - Q_n(\hbbeta) =  \frac{1}{n}\left\{\by\t\bX\bdelta - \bone\t \left[\bb(\bX\hbbeta)-\bb(\bX\bbeta_0)\right]\right\}  + \|p_{\lambda }(\bbeta_0)\|_1 - \|p_{\lambda }(\hbbeta)\|_1.
\end{equation}
To analyze the nonlinear term $\bone\t [\bb(\bX\hbbeta)-\bb(\bX\bbeta_0)]$, we do a second-order Taylor expansion of the function $\bone\t [\bb(\bX \bbeta)-\bb(\bX\bbeta_0)]$ around $\bbeta_0$ and retain the Lagrange remainder term, which gives
\begin{equation} \label{f002}
\bone\t [\bb(\bX\hbbeta)-\bb(\bX\bbeta_0)] = \left\{\bb'(\bX\bbeta_0)\right\}\t\bX\bdelta + \frac{1}{2} \bdelta\t\bX\t\bH(\tbbeta)\bX\bdelta,
\end{equation}
where $\bH(\tbbeta) = \text{diag}\{\bb''(\bX\tbbeta)\}$ is a diagonal matrix with $\tbbeta \in \mathbb{R}^p$ lying on the line segment connecting $\bbeta_0$ and $\hbbeta$. Thus combining inequality (\ref{f001}) with representation (\ref{f002}) yields
\begin{equation} \label{f004}
0\leq n^{-1} \left[ \bveps\t\bX\bdelta - \frac{1}{2} \bdelta\t\bX\t\bH(\tbbeta)\bX\bdelta \right] + \|p_{\lambda }(\bbeta_0)\|_1 - \|p_{\lambda }(\hbbeta)\|_1,
\end{equation}
where $\bveps = \by - E \by = \by - \bb'(\bX\bbeta_0)$ denotes the $n$-dimensional error vector in the GLM. We observe that the first term on the right hand side of (\ref{f004}) resembles the corresponding one in the case of linear model.

A rearrangement of the above inequality (\ref{f004}) gives
\begin{align}\label{e001}
(2n)^{-1}\bdelta\t\bX\t\bH(\tbbeta)\bX\bdelta \leq n^{-1}\bveps\t\bX\bdelta + \|p_{\lambda }(\bbeta_0)\|_1 - \|p_{\lambda }(\hbbeta)\|_1.
\end{align}
It follows from $\bbeta_0, \hbbeta \in \mathcal{B}_\tau$ that
$\|\bdelta\|_0 \leq \|\bbeta_0\|_0+\|\hbbeta\|_0 < \kappa_c$.  Thus, by Condition \ref{cond2} and the robust spark definition, the left hand side of (\ref{e001}) can be bounded as
\begin{align*}
n^{-1}\bdelta\t\bX\t\bH(\tbbeta)\bX\bdelta \geq c_2 n^{-1}\|\bX\bdelta\|_2^2 = c_2 n^{-1}\|\bX_{\supp(\sbdelta)}\bdelta_{\supp(\sbdelta)}\|_2^2 \geq c c_2 \|\bdelta\|_2^2.
\end{align*}
On the other hand, the first term on the right hand side of (\ref{e001}) can be bounded as
\[
|n^{-1}\bveps\t\bX\bdelta| \leq \|n^{-1}\bX\t\bveps\|_{\infty}\|\bdelta\|_1.
\]
The concavity of the penalty function $p_\lambda(t)$ assumed in Condition \ref{cond3} entails that $p_\lambda'(t)$ is decreasing in $t$, which leads to $p_{\lambda }'(t) \leq p'_{\lambda }(0+) =c_3\lambda $ for any $t \geq 0$. Thus, it follows from the mean value theorem and triangular inequality that
\begin{align}\label{e025}
\big|\|p_{\lambda }(\bbeta_0)\|_1 - \|p_{\lambda }(\hbbeta)\|_1\big| = \big|\sum\nolimits_{j=1}^p p'_{\lambda }(t_j) \big( |\beta_{0,j}|-|\hbeta_j|\big)\big| \leq c_3\lambda  \|\bdelta\|_1,
\end{align}
where $t_j$ lies between $|\beta_{0,j}|$ and $|\hbeta_j|$ for $j=1,\cdots, p$ and $\hbbeta = (\hbeta_1, \cdots, \hbeta_p)\t$. Combining the above three results with (\ref{e001}) completes the proof.

\subsection{Proof of Theorem \ref{Thm1}}  \label{SecB.4}
We first show the existence of the global minimizer and then prove the bounds under different losses.

\medskip

{\noindent \bf Existence of global minimizer:} Since the negative log-likelihood function is smooth by Condition \ref{cond2} and the penalty function is continuous by Condition \ref{cond3}, we see that the objective function $Q_n(\bbeta)$ is continuous. Let $R$ be any subspace of $\mathbb{R}^p$ with dimension less than $\kappa_c/2$ and denote by $L(\bbeta) = -n^{-1}\{\by\t\bX\bbeta - \bone\t\bb(\bX\bbeta)\}$ the negative log-likelihood function. It follows from Condition \ref{cond2} and the definition of the robust spark that $L(\bbeta)$ is strictly convex on $R$ and its Hessian matrix has the smallest eigenvalue bounded from below by $c_2 c^2$. Applying the second-order Taylor expansion around $\bzero$ with the Lagrange remainder term shows that $L(\bbeta)$ is bounded from below by $\widetilde{L}(\bbeta) = -n^{-1}\by\t\bX\bbeta + b(0) + n^{-1}\{\bb'(\bzero)\}\t \bX \bbeta + c_2 c^2 \|\bbeta\|_2^2$ for any $\bbeta \in R$, with $L(\bzero) = \widetilde{L}(\bzero) = b(0)$. This entails that there exists some sufficiently large positive number $C$, which is independent of the subspace $R$, such that
\[ L(\bbeta) \geq \widetilde{L}(\bbeta) > \widetilde{L}(\bzero) = L(\bzero) \]
for any $\bbeta \in R$ with $\|\bbeta\|_2 > C$. Thus the global minimizer of $Q_n(\bbeta) = L(\bbeta) + \|p_\lambda(\bbeta)\|_1$ on the thresholded parameter space $\Btau$ must lie in $T = \Btau \cap \{\bbeta \in \mathbb{R}^p: \|\bbeta\|_2 \leq C\}$. In view of (\ref{eq:Bset}), $\Btau$ is a closed set and thus $T$ is a compact set. Therefore, the existence of the global minimizer of $Q_n(\bbeta)$ over $\Btau$ is guaranteed by its continuity.

\medskip

{\noindent \bf False signs:} We use the induction method to prove the result. Let $\bdelta = (\delta_1 ,\cdots, \delta_p)\t = \hbbeta -\bbeta_0$. Since $\|\bdelta\|_0 \leq \|\bbeta_0\|_0 + \|\hbbeta\|_0<\kappa_c$, by the Cauchy-Schwarz inequality, we have $\|\bdelta\|_1 \leq \sqrt{\kappa_c}\|\bdelta\|_2$. Hence, it follows from Lemma \ref{Lem1} that conditional on the event $\mathcal{E}$,
\begin{equation}\label{e004}
\|\bdelta\|_2^2 \leq C\lambda \|\bdelta\|_1\leq C\lambda \sqrt{\kappa_c}\|\bdelta\|_2.
\end{equation}
Solving for $\|\bdelta\|_2$ yields
\begin{equation}\label{e006}
\|\bdelta\|_2 \leq C\lambda  \sqrt{\kappa_c}.
\end{equation}
 On the other hand, it follows from $\hbbeta, \bbeta_0 \in \mathcal{B}_\tau$ that $\|\bdelta\|_2 \geq \{\FS(\hbbeta)\}^{1/2}\tau$. This together with (\ref{e006}) ensures that
\begin{equation*}
\FS(\hbbeta) \leq  C(\lambda /\tau)^2\kappa_c.
\end{equation*}
Thus we have $\|\bdelta\|_0 \leq \|\bbeta_0\|_0 + \FS(\hbbeta) \leq s + C(\lambda /\tau)^2\kappa_c$.
So the upper bound $s + C(\lambda /\tau)^2\kappa_c$ plays the same role as $\kappa_c$ in (\ref{e006}). Repeating the above derivations with $\kappa_c$ replaced with $s +C(\lambda /\tau)^2\kappa_c $ and by induction, we have
$\FS(\hbbeta)\leq  Cs\lambda ^2\tau^{-2}/(1-C\lambda ^2\tau^{-2})$ conditional on $\mathcal{E}$, which completes the proof of the result on false signs.

\medskip

{\noindent\bf Estimation losses:} We first prove the inequalities under the $L_1$- and $L_2$-norms, and then use H\"{o}lder's inequality to prove the general result under the $L_q$-norm with $q\in (1,2)$. The  result on $L_\infty$-norm follows immediately from the $L_2$-norm result. By default, all arguments are conditioning on $\mathcal{E}$ in Condition \ref{cond1}, which holds with probability at least $1-O(p^{-c_1})$.

Since $\|\bdelta\|_0 \leq \|\bbeta_0\|_0 + \FS(\hbbeta)$, by the Cauchy-Schwarz inequality and the result on $\FS(\hbbeta)$ proved above we have
\[
\|\bdelta\|_1 \leq \|\bdelta\|_0^{1/2}\|\bdelta\|_2 \leq C\{s/(1-C\lambda ^2\tau^{-2})\}^{1/2}\|\bdelta\|_2.
\]
This together with the first inequality in (\ref{e004}) entails that
\begin{equation}\label{e020}
\|\bdelta\|_2 \leq C\lambda \{s/(1-C\lambda ^2\tau^{-2})\}^{1/2}.
\end{equation}
Combining the above two inequalities we obtain
\begin{equation}\label{e007}
\|\bdelta\|_1 \leq C\lambda s/(1-C\lambda ^2\tau^{-2}).
\end{equation}
Finally, for $q\in (1,2)$, applying H\"{o}lder's inequality and in view of (\ref{e020}) and (\ref{e007}), we have
\begin{align}\label{e029}
\|\bdelta\|_q = \big(\sum\nolimits_{j=1}^p|\delta_j|^{2-q}|\delta_j|^{2q-2}\big)^{1/q}\leq \|\bdelta\|_1^{(2-q)/q}\|\bdelta\|_2^{2(q-1)/q} \leq C\lambda \big\{s/(1-C\lambda ^2\tau^{-2})\big\}^{1/q}.
\end{align}
The oracle inequality on $\|\bdelta\|_\infty$ follows immediately from $\|\bdelta\|_\infty \leq \|\bdelta\|_2$ and (\ref{e020}). This completes the proof for the estimation losses.

\medskip
{\noindent \bf Prediction loss:} The inequality for this loss follows immediately from plugging (\ref{e007}) into Lemma \ref{Lem1} and using Condition \ref{cond1}, which concludes the proof.

\subsection{Proof of Theorem \ref{Thm2}}  \label{SecB.5}
Define an event $\mathcal{E}_1 = \mathcal{E}\cap\mathcal{E}_0$ with $\mathcal{E}$ and $\mathcal{E}_0$ defined in (\ref{eq:Eset}). We will prove that all results in Theorem \ref{Thm2} hold simultaneously on the event $\mathcal{E}_1$. Then Theorem \ref{Thm2} follows immediately from Condition \ref{cond1}. By default, all arguments in this proof are conditioning on $\mathcal{E}_1$.

\medskip

{\noindent \bf Sign consistency:}
Denote by $\alpha = \supp(\hbbeta)$ and $\alpha_0 = \supp(\bbeta_0)$. We use the method of proof by contradiction to show that we must have $\alpha = \alpha_0$. Let $\hbbeta^*$ be the oracle-assisted maximum likelihood estimator. We make use of the following decomposition:
\begin{align*}
 Q_n(\hbbeta) - Q_n(\hbbeta^*) = I_1 + I_2,
\end{align*}
where
$I_1 = -n^{-1}\by\t\bX(\hbbeta - \hbbeta^*) + n^{-1}\bone\t[\bb(\bX\hbbeta) - \bb(\bX\hbbeta^*)]$ and
$I_2 = \|p_{\lambda }(\hbbeta)\|_1 - \|p_{\lambda }(\hbbeta^*)\|_1$.
We will prove that $\hbbeta^* \in \mathcal{B}_\tau$, and that if $\alpha \neq \alpha_0$, then
\begin{align}
&I_1  \geq cc_2\tau^2/4, \label{e011}\\
&|I_2| \leq o(\tau^2). \label{e017}
\end{align}
Combining the above results, we have $Q_n(\hbbeta) - Q_n(\hbbeta^*) > 0$ for sufficiently large $n$, which contradicts with $\hbbeta$ being a global minimizer in $\mathcal{B}_\tau$, and thus we must have $\alpha = \alpha_0$. On the other hand, since $\lambda = o(\tau/\sqrt{s})$, Theorem \ref{Thm1} ensures that for large enough $n$,
\begin{equation}\label{e002}
\|\hbbeta - \bbeta_0\|_\infty \leq C\sqrt{s}\lambda = o(\tau).
\end{equation}
This together with $\alpha = \alpha_0$ and $\bbeta_0, \hbbeta \in \Btau$ entails that conditioning on $\mathcal{E}_1$, $\sgn(\hbbeta) = \sgn(\bbeta_0)$. Thus, the sign consistency result follows easily from Condition \ref{cond1}.

We first proceed to prove (\ref{e017}). By definition, $\supp(\hbbeta^*) = \alpha_0$ and $\hbbeta^*_{\alpha_0}= (\hbeta_1^*,\cdots, \hbeta_s^*)\t$ minimizes the negative log-likelihood function $Q_n^*(\bbeta_{\alpha_0}) = -\by\t\bX_{\alpha_0}\bbeta_{\alpha_0} + \bone\t\bb(\bX_{\alpha_0}\bbeta_{\alpha_0})$ with $\bbeta_{\alpha_0} \in \mathbb{R}^{s}$. Thus, $\hbbeta_{\alpha_0}^*$ is a critical point of $Q_n^*(\bbeta_{\alpha_0})$ and satisfies
\begin{align}\label{e014}
-\bX\t_{\alpha_0}\left[\by -\bb'(\bX_{\alpha_0}\hbbeta_{\alpha_0}^*)\right] = \bzero.
\end{align}
Plugging the true model $\by = \bb'(\bX_{\alpha_0}\bbeta_{0,\alpha_0}) + \bveps$ into (\ref{e014}) and applying the mean value theorem componentwise, we have
\begin{equation} \label{f005}
-\bX_{\alpha_0}\t\bveps + \bX_{\alpha_0}\t \bH(\tbbeta_1, \cdots, \tbbeta_n)\bX_{\alpha_0}(\hbbeta_{\alpha_0}^* - \bbeta_{0,\alpha_0})= \bzero,
\end{equation}
where $\bH(\tbbeta_1, \cdots, \tbbeta_n) = \text{diag}\{b''(\bx_1\t\tbbeta_1), \cdots, b''(\bx_n\t\tbbeta_n)\}$ with each $\tbbeta_i = (\tbeta_{i,1}, \cdots, \tbeta_{i,p})\t$ lying on the line segment connecting $\bbeta_{0}$ and $\hbbeta^*$. The above equation can be rewritten as
\begin{align}\label{e050}
\hbbeta^*_{\alpha_0} - \bbeta_{0,\alpha_0}=\left\{\bX\t_{\alpha_0}\bH(\tbbeta_1, \cdots, \tbbeta_n)\bX_{\alpha_0}\right\}^{-1}\bX_{\alpha_0}\t\bveps.
\end{align}
Therefore, by Condition \ref{cond2},  we obtain that conditioning on $\mathcal{E}$,
\begin{align}\label{e064}
\|\hbbeta^*_{\alpha_0} - \bbeta_{0,\alpha_0}\|_2 \leq  C\|\bX_{\alpha_0}\t\bveps\|_2/n\leq C\sqrt{s}\|\bX_{\alpha_0}\t\bveps\|_\infty/n\leq C\sqrt{s}\lambda.
\end{align}
This together with the assumptions $\min_{j\leq s}|\beta_{0,j}|\geq 2\tau$  and $\sqrt{s}\lambda = o(\tau)$ entails that
\begin{equation*}
\min_{1\leq j\leq s}|\hbeta_j^*|\geq 2\tau - C\sqrt{s}\lambda >\tau \text{ and thus }\hbbeta^*\in \mathcal{B}_\tau.
\end{equation*}
Similarly to (\ref{e025}) and by Theorem \ref{Thm1} and (\ref{e064}), we can prove
\begin{align}\label{e015}
\nonumber\big|I_2 \big|&\leq p_{\lambda }'(0+)\|\hbbeta - \hbbeta^*\|_1\leq c_3\lambda (\|\hbbeta - \bbeta_0\|_1 + \|\hbbeta^* - \bbeta_0\|_1)\\
& \leq  c_3\lambda \big(\|\hbbeta - \bbeta_0\|_1 + \sqrt{s}\|\hbbeta^* - \bbeta_0\|_2\big) \leq Cs\lambda ^2 = o(\tau^2).
\end{align}
This completes the proof of (\ref{e017}) and $\hbbeta^*\in \mathcal{B}_\tau$.

It remains to prove (\ref{e011}). Applying the second-order Taylor expansion around $\hbbeta^*$ with the Lagrange remainder term, $I_1$ can be decomposed as
\begin{align}\label{e068}
 I_1& = -\frac{1}{n}[\by - \bb'(\bX\hbbeta^*)]\t\bX(\hbbeta - \hbbeta^*) + \frac{1}{2 n}(\hbbeta - \hbbeta^*)\t\bX\t\bH(\tbbeta^*)\bX(\hbbeta - \hbbeta^*)  \equiv I_{1,1} + I_{1,2},
\end{align}
where $\bH(\tbbeta^*) = \diag\{\bb''(\bX\tbbeta^*)\}$ with $\tbbeta^*$ lying on the line segment connecting $\hbbeta^*$ and $\hbbeta$.  It follows from $\hbbeta, \hbbeta^* \in \Btau$ that  $\|\hbbeta - \hbbeta^*\|_0  <\kappa_c$. Thus, by Condition \ref{cond2}, the robust spark definition, and $\hbbeta, \hbbeta^* \in \mathcal{B}_\tau$, we have
\begin{align}\label{e069}
I_{1,2}\geq \frac{1}{2}c_2c\|\hbbeta - \hbbeta^*\|_2^2 \geq \frac{1}{2}c_2c\tau(\|\hbbeta_{\alpha\setminus\alpha_0}\|_1 + \|\hbbeta^*_{\alpha_0\setminus\alpha}\|_1).
\end{align}

We now consider the term $I_{1,1}$ in (\ref{e068}). By (\ref{e014}), we have
\begin{align}\label{e049}
\nonumber I_{1,1}&=-n^{-1}[\by - \bb'(\bX_{\alpha_0}\hbbeta^*_{\alpha_0})]\t\bX_{\alpha_0}(\hbbeta_{\alpha_0}-\hbbeta_{\alpha_0}^*)- n^{-1}[\by - \bb'(\bX\hbbeta^*)]\t\bX_{\alpha\setminus\alpha_0}\hbbeta_{\alpha\setminus\alpha_0}\\
& =-n^{-1}[\by - \bb'(\bX\hbbeta^*)]\t\bX_{\alpha\setminus\alpha_0}\hbbeta_{\alpha\setminus\alpha_0}.
\end{align}
Plugging $\by = \bb'(\bX\bbeta_{0}) + \bveps$  into (\ref{e049}) and by the mean value theorem, we have
\begin{align}\label{e052}
I_{1,1} = -n^{-1}\bveps\t\bX_{\alpha\setminus\alpha_0}\hbbeta_{\alpha\setminus\alpha_0} + n^{-1}(\hbbeta^*_{\alpha_0}-\bbeta_{0,\alpha_0})\t\bX_{\alpha_0}\t \bH(\tbbeta)\bX_{\alpha\setminus\alpha_0} \hbbeta_{\alpha\setminus\alpha_0},
\end{align}
where $\bH(\tbbeta) = \diag\{\bb''(\bX\tbbeta)\}$ with $\tbbeta$ lying on the line segment connecting $\bbeta_0$ and $\hbbeta^*$. Conditioning on $\mathcal{E}$, the first term of (\ref{e052}) can be bounded as
\begin{equation}\label{e066}
n^{-1}|\bveps\t\bX_{\alpha\setminus\alpha_0}\hbbeta_{\alpha\setminus\alpha_0}| \leq\|n^{-1}\bX_{\alpha\setminus\alpha_0}\t \bveps\|_{\infty}\|\hbbeta_{\alpha\setminus\alpha_0}\|_1  \leq \lambda \|\hbbeta_{\alpha\setminus\alpha_0}\|_1.
\end{equation}

Next we study the second term of (\ref{e052}). We will make use of (\ref{e050}). By the Cauchy-Schwarz inequality, Condition \ref{cond2}, and the robust spark definition, we have
\begin{align}
\nonumber \|\hbbeta^*_{\alpha_0} - \bbeta_{0,\alpha_0}\|_\infty &\leq \|\hbbeta^*_{\alpha_0} - \bbeta_{0,\alpha_0}\|_2 \leq \|\big\{n^{-1}\bX\t_{\alpha_0}\bH(\tbbeta_1,\cdots,\tbbeta_n)\bX_{\alpha_0}\big\}^{-1}\|_2\| n^{-1}\bX_{\alpha_0}\t\bveps\|_2 \\
&\leq C \sqrt{s}\| n^{-1}\bX_{\alpha_0}\t\bveps\|_\infty  \leq C\sqrt{s(\log n)/n}. \label{e079}
\end{align}
Recall that $\tbbeta$ defined in (\ref{e052}) lies on the line segment connecting $\bbeta_0$ and $\hbbeta^*$. Thus, by (\ref{e064}), we have $\|\tbbeta - \bbeta_0\|_2 \leq \|\hbbeta^*- \bbeta_0\|_2 \leq C\sqrt{s}\lambda$, which ensures that $\tbbeta \in \mathcal{B}_1^*$ with $\mathcal{B}_1^*$ defined in (\ref{e047}).
Since Theorem \ref{Thm1} ensures that $|\alpha\setminus \alpha_0| \leq \FS(\hbbeta) \leq s$ for large enough $n$, it follows from the above inequality (\ref{e079}) that the second term of (\ref{e052}) can be bounded as
\begin{align}\label{e067}
\nonumber &n^{-1}|(\hbbeta^*_{\alpha_0}-\bbeta_{0,\alpha_0})\t\bX_{\alpha_0}\t\bH(\tbbeta)\bX_{\alpha\setminus\alpha_0}\hbbeta_{\alpha\setminus\alpha_0}| \\ &\leq\|n^{-1}\bX_{\alpha\setminus\alpha_0}\t\bH(\tbbeta)\bX_{\alpha_0}\|_{\infty}\|\hbbeta^*_{\alpha_0}-\bbeta_{0,\alpha_0}\|_\infty \|\hbbeta_{\alpha\setminus\alpha_0}\|_1 \leq C\gamma_n\sqrt{s(\log n)/n}\|\hbbeta_{\alpha\setminus\alpha_0}\|_1.
\end{align}
Combining (\ref{e066}) and (\ref{e067}) and in view of (\ref{e052}), we obtain that
\begin{equation*}
|I_{1,1}| \leq \left[\lambda  +C\gamma_n\sqrt{s(\log n)/n} \right]\|\hbbeta_{\alpha\setminus\alpha_0}\|_1.
\end{equation*}
 This together with (\ref{e068}), (\ref{e069}), $\hbbeta^*, \hbbeta \in \mathcal{B}_\tau$, and the assumption $\tau \gg \max\{\lambda , \gamma_n\sqrt{s(\log n)/n}\}$ ensures that if $\alpha \neq \alpha_0$, then for large enough $n$,
\begin{align*}
I_1&\geq \frac{1}{2}cc_2\tau \|\hbbeta^*_{\alpha_0\setminus\alpha}\|_1 +
\left[\frac{1}{2}cc_2\tau-\lambda  -C\gamma_n\sqrt{s(\log n)/n}\right]\|\hbbeta_{\alpha\setminus\alpha_0}\|_1 \geq cc_2\tau^2/4,
\end{align*}
which proves (\ref{e011}) and completes the proof of sign consistency.

\medskip
{\noindent \bf Estimation losses:} We first prove for the $L_\infty$-estimation loss. By (\ref{e002}) and the sign consistency proved above, we have $\supp(\hbbeta)=\alpha_0$ and $\min_{1\leq j\leq s}|\hbeta_j| \geq \min_{1 \leq j\leq s}|\beta_{0,j}| - o(\tau) > \tau$ for large enough $n$. Thus, $\hbbeta$ is an interior point of $\mathcal{B}_\tau$. Since $\hbbeta$ is the global minimizer, it follows that $\hbbeta_{\alpha_0}$ is a critical point of $Q_n(\bbeta_{\alpha_0}, \bzero)$ and satisfies
\begin{align}\label{e012}
-\bX_{\alpha_0}\t\by + \bX_{\alpha_0}\t\bb'(\bX\hbbeta) + n\bar{p}_{\lambda }(\hbbeta_{\alpha_0}) = \bzero,
\end{align}
where $\bar{p}_{\lambda }(\hbbeta_{\alpha_0})$ is an $s$-dimensional vector with components $p_{\lambda }'(|\hbeta_j|)\sgn(\hbeta_j)$ for $j \in \alpha_0$. Similarly to (\ref{f005}), plugging $\by = \bb'(\bX\bbeta_0) + \bveps$ into (\ref{e012}) and applying the mean value theorem componentwise,  we have
\begin{align}\label{e003}
n\bar{p}_{\lambda }(\hbbeta_{\alpha_0}) & = -\bX_{\alpha_0}\t\bH(\tbbeta_1, \cdots,\tbbeta_n)\bX_{\alpha_0}(\hbbeta_{\alpha_0} - \bbeta_{0,\alpha_0}) + \bX_{\alpha_0}\t\bveps,
\end{align}
where $\bH(\tbbeta_1, \cdots ,\tbbeta_n)$ is defined similarly as in (\ref{f005}) with each $\tbbeta_i$ lying on the line segment connecting $\bbeta_0$ and $\hbbeta$. Thus, for each $1\leq i \leq n$, $\supp(\tbbeta_i) = \alpha_0$, and by Theorem \ref{Thm1}, $\|\tbbeta_i - \bbeta_0\|_2 \leq \|\hbbeta - \bbeta_0\|_2 \leq 2C\lambda \sqrt{s}$ for large enough $n$, which ensures that $\tbbeta_i \in \mathcal{B}_1^*$ with $\mathcal{B}_1^*$ defined in (\ref{e047}). So (\ref{e003}) can be rewritten as
\begin{align}\label{e048}
\nonumber \hbbeta_{\alpha_0} - \bbeta_{0,\alpha_0} &= \left[\bX_{\alpha_0}\t \bH(\tbbeta_1, \cdots ,\tbbeta_n) \bX_{\alpha_0}\right]^{-1}\bX_{\alpha_0}\t\bveps - n \left[\bX_{\alpha_0}\t \bH(\tbbeta_1, \cdots ,\tbbeta_n) \bX_{\alpha_0}\right]^{-1} \bar p_{\lambda}(\hbbeta_{\alpha_0}) \\
&\equiv I_1(\alpha_0) + I_2(\alpha_0).
\end{align}

We first study $I_1(\alpha_0)$. Since $\tbbeta_i \in \mathcal{B}_1^*$, by the Cauchy-Schwarz inequality,
\begin{align}\label{e008}
 \|I_1(\alpha_0)\|_\infty \leq \left\|\left[n^{-1} \bX_{\alpha_0}\t \bH(\tbbeta_1, \cdots ,\tbbeta_n) \bX_{\alpha_0}\right]^{-1}\right\|_{\infty} \|n^{-1}\bX_{\alpha_0}\t\bveps\|_\infty \leq C\gamma_n^* \sqrt{(\log n)/n}.
\end{align}
Next we study $I_2(\alpha_0)$. Similarly, since $p_{\lambda}(t)$ is a concave penalty, $\supp(\hbbeta)=\alpha_0$, and $\hbbeta \in \mathcal{B}_\tau$,  we can prove that
\begin{align}\label{e009}
\|I_2(\alpha_0)\|_\infty \leq  \left\|\left[n^{-1} \bX_{\alpha_0}\t \bH(\tbbeta_1, \cdots ,\tbbeta_n) \bX_{\alpha_0}\right]^{-1}\right\|_{\infty} \|\bar p_{\lambda }(\hbbeta_{\alpha_0})\|_{\infty}\leq  \gamma_n^*p'_{\lambda }(\tau).
\end{align}
Therefore, if $p'_{\lambda }(\tau) =  O\big\{ \sqrt{(\log n)/n} \big\}$, then conditioning on $\mathcal{E}_1$,
\[
\|\hbbeta_{\alpha_0} - \bbeta_{0,\alpha_0}\|_{\infty} \leq \|I_1(\alpha_0)\|_\infty  + \|I_2(\alpha_0)\|_\infty \leq C\gamma_{n}^*\sqrt{(\log n)/n},
\]
which completes the proof of the oracle inequality under the $L_{\infty}$-estimation loss.

We now study the $L_q$-estimation loss with $q\in [1,2]$. Similarly as in Theorem \ref{Thm1}, we first prove results under the $L_1$- and $L_2$-norms, and then use H\"{o}lder's inequality to prove the general results. Since $\supp(\hbbeta) = \alpha_0$, $p'_{\lambda }(\tau) = O\big\{\sqrt{(\log n)/n}\big\}$, and $p'_{\lambda}(t)$ is decreasing in $t\in (0,\infty)$, inequality (\ref{e025}) in the proof of Lemma \ref{Lem1} can be bounded as
\[
\Big|\|p_{\lambda }(\bbeta_0)\|_1 - \|p_{\lambda }(\hbbeta)\|_1\Big| = \Big|\sum\nolimits_{j\in\alpha_0}p'_{\lambda }(t_j) \big( |\beta_{0,j}|-|\hbeta_j|\big)\Big| \leq p'_{\lambda }(\tau) \|\bdelta\|_1 \leq C\sqrt{(\log n)/n}\|\bdelta\|_1,
\]
where the second step is because $t_j$ is between $|\beta_{0,j}|$ and $|\hbeta_j|$ and thus $t_j \geq \tau$ for each $j\in\alpha_0$. Using similar proof as in Lemma \ref{Lem1} and the above inequality, we obtain that conditioning on $\mathcal{E}_1$,
\begin{align}\label{e071}
\|\bdelta\|_2^2 \leq c^{-1}n^{-1}\|\bX\bdelta\|_2^2\leq C\sqrt{(\log n)/n}\|\bdelta\|_1.
\end{align}
Since the sign consistency proved above ensures $\|\bdelta\|_1 \leq \sqrt{s}\|\bdelta\|_2$, it follows from (\ref{e071}) that
\begin{align}\label{e010}
 \|\bdelta\|_2 \leq \sqrt{s}\|\bdelta\|_2^2/\|\bdelta\|_1\leq C\sqrt{s(\log n)/n} \ \text{ and } \ \|\bdelta\|_1 \leq \sqrt{s}\|\bdelta\|_2 \leq Cs\sqrt{(\log n)/n}.
\end{align}
The oracle inequalities under the $L_q$-estimation loss with $q\in (1,2)$ follow immediately from H\"{o}lder's inequality and (\ref{e010}), as in (\ref{e029}). Thus, the results on estimation losses are proved.

\medskip
{\noindent \bf Prediction loss:} Since $E \bY =\bb'(\bX\bbeta_0)$, it follows from the second-order Taylor expansion around $\bbeta_0$ with the Lagrange remainder term that
\begin{align}\label{e031}
D(\hbbeta) & = \frac{1}{2}\bdelta\t\bX\t\bH(\tbbeta)\bX\bdelta,
\end{align}
where $\bH(\tbbeta) = \text{diag}\{\bb''(\bX\tbbeta)\}$ with $\tbbeta$ lying on the line segment connecting $\bbeta_0$ and $\hbbeta$.
Since $\supp(\hbbeta) = \alpha_0$, it follows from Condition \ref{cond2}, (\ref{e071}), and (\ref{e010})that
\begin{align*}
&\bdelta\t\bX\t\bH(\tbbeta)\bX\bdelta \leq c_2^{-1}\|\bX\bdelta\|_2^2 \leq C\sqrt{n(\log n)}\|\bdelta\|_1 \leq Cs(\log n).
\end{align*}
Thus, combining the above inequality with (\ref{e031}) completes the proof.

\subsection{Proof of Theorem \ref{Thm3}}  \label{SecB.6}
Define $\mathcal{E}_1 = \mathcal{E}\cap \mathcal{E}_0$ as in the proof of Theorem \ref{Thm2}. Then all results in Theorem \ref{Thm2} hold simultaneously on the event $\mathcal{E}_1$, which satisfies $P(\mathcal{E}_1^c) = O(n^{-c_1})$ by Condition \ref{cond1}. Denote by $\bdelta = \hbbeta - \bbeta_0$.

\medskip
{\noindent \bf Estimation risks}: Similarly as in Theorem \ref{Thm2}, we first prove the results under the $L_2$- and $L_1$-losses, and then use H\"{o}lder's inequality to prove the general result under the $L_q$-loss with $q\in (1,2)$. We first show that
\begin{equation}\label{e028}
E \|\bdelta\|_2^2 \leq Cs(\log n)/n.
\end{equation}
The key is to prove the following three inequalities:
\begin{align}
& E\left\{\|\bdelta\|_2^21_{\mathcal{E}_1}\right\}  \leq Cs(\log n)/n, \label{e061}\\
&  E\left\{\|\bdelta\|_2^21_{\mathcal{E}^c}\right\} =  O(p^{-c_1/2}\kappa_c), \label{e062}\\
&  E\left\{\|\bdelta\|_2^21_{\mathcal{E}_0^c\cap \mathcal{E}}\right\} = O(\lambda sn^{-c_1}). \label{e063}
\end{align}
Since $\mathcal{E}_1^c = \mathcal{E}^c \cup(\mathcal{E}_0^c \cap \mathcal{E})$, $c_1$ can be chosen arbitrarily large, and $\kappa_c \leq n + 1$, the inequality (\ref{e028}) follows immediately by combining (\ref{e061})--(\ref{e063}).

We first proceed to prove (\ref{e061}). By (\ref{e010}) in the proof of Theorem \ref{Thm2}, (\ref{e061}) can be proved as follows:
\[
E[\|\bdelta\|_21_{\mathcal{E}_1}] \leq E[C\sqrt{s(\log n)/n}1_{\mathcal{E}_1}] \leq C\sqrt{s(\log n)/n}.
\]

Next we prove (\ref{e062}). By the Cauchy-Schwarz inequality and Condition \ref{cond1}, we have
\begin{align}\label{e032}
\nonumber & E[\|n^{-1}\bX\t\bveps\|_\infty^21_{\mathcal{E}^c}]\leq n^{-2}\{E[\|\bX\t\bveps\|_\infty^4] P(\mathcal{E}^c)\}^{1/2}\\
&\leq O(p^{-c_1/2})n^{-2} \left\{E \left[\max_{1\leq j\leq p}\|\widetilde{\bx}_j\|_2^4\|\bveps\|_2^4\right]\right\}^{1/2}=O(p^{-c_1/2}),
\end{align}
where the last step is because of $\|\widetilde{\bx}_j\|_2 = \sqrt{n}$ and the assumption $\max_{1\leq i\leq n}E \veps_i^4\leq C$.
Similarly, we can prove that
\begin{align}\label{e035}
E[\|n^{-1}\bX\t\bveps\|_\infty1_{\mathcal{E}^c}]=O(p^{-c_1/2}) \ \text{ and } \
 E[\|n^{-1}\bX\t\bveps\|_\infty^21_{\mathcal{E}_0^c}]=O(n^{-c_1/2}).
\end{align}
Since $\|\bdelta\|_0 \leq \|\bbeta_0\|_0 + \|\hbbeta\|_0 < \kappa_c$, it follows that  $\|\bdelta\|_1 \leq \sqrt{\kappa_c}\|\bdelta\|_2$. This together with (\ref{e026}) in Lemma \ref{Lem1} yields
\begin{align}\label{e040}
\|\bdelta\|_21_{\mathcal{E}^c} \leq C\sqrt{\kappa_c}\big(\|n^{-1}\bX\t\bveps\|_\infty + \lambda \big)1_{\mathcal{E}^c}.
\end{align}
Thus, by Condition \ref{cond1} and (\ref{e032}), the inequality (\ref{e062}) is proved as follows:
\begin{align}
 E[\|\bdelta\|_2^21_{\mathcal{E}^c}] \leq C\kappa_cE[\|n^{-1}\bX\t\bveps\|_\infty^21_{\mathcal{E}^c}] + C\kappa_c\lambda ^2P(\mathcal{E}^c) 
 = O(p^{-c_1/2}\kappa_c). \label{e036}
\end{align}

Finally we prove (\ref{e063}). To this end, we first prove
\begin{align}\label{e039}
E[\|\bdelta\|_11_{\mathcal{E}_0^c\cap \mathcal{E}} ] = O(s\lambda  n^{-c_1}).
\end{align}
Then by Lemma \ref{Lem1} and the definition of $\mathcal{E}$, (\ref{e063}) can be proved as follows:
\begin{align}\label{e041}
&E[\|\bdelta\|_2^21_{\mathcal{E}_0^c\cap \mathcal{E}}] \leq  C\lambda  E[\|\bdelta\|_11_{\mathcal{E}_0^c\cap \mathcal{E}}] = O(\lambda ^2sn^{-c_1}).
\end{align}

It remains to prove (\ref{e039}). We first study $E[\|\bdelta\|_11_{\mathcal{E}_0^c\cap\mathcal{E}}]$ by decomposing it into two terms:
\begin{equation}\label{e024}
E[\|\bdelta\|_11_{\mathcal{E}_0^c\cap \mathcal{E}}] = E[\|\bdelta_{\alpha_0^c}\|_11_{\mathcal{E}_0^c\cap \mathcal{E}}] + E[\|\bdelta_{\alpha_0}\|_11_{\mathcal{E}_0^c\cap \mathcal{E}}].
\end{equation}
We now consider the first term on the right hand side of (\ref{e024}).
Since $\hbbeta \in \Btau$, it follows that  $\|\bdelta\|_2^2 = \|\bdelta_{\alpha_0}\|_2^2 + \|\bdelta_{\alpha_0^c}\|_2^2 \geq s^{-1}\|\bdelta_{\alpha_0}\|_1^2 + \tau \|\bdelta_{\alpha_0^c}\|_1$. Thus, by Lemma \ref{Lem1} we have conditioning on $\mathcal{E}_0^c\cap \mathcal{E}$,
\[
s^{-1}\|\bdelta_{\alpha_0}\|_1^2 + \tau \|\bdelta_{\alpha_0^c}\|_1 \leq \|\bdelta\|_2^2  \leq C\lambda \|\bdelta\|_1= C \lambda (\|\bdelta_{\alpha_0}\|_1 + \|\bdelta_{\alpha_0^c}\|_1 ).
\]
A rearrangement of the above inequality yields
\begin{align}\label{e037}
 & \big(\|\bdelta_{\alpha_0}\|_1 - Cs\lambda\big)^2 \leq s \big(C\lambda - \tau\big)\|\bdelta_{\alpha_0^c}\|_1 + Cs^2\lambda^2.
\end{align}
Since the left hand side of (\ref{e037}) is always nonnegative and $\lambda = o(\tau/\sqrt{s})$, we have
$
\|\bdelta_{\alpha_0^c}\|_11_{\mathcal{E}_0^c\cap \mathcal{E}} \leq C\tau^{-1}s\lambda^21_{\mathcal{E}_0^c\cap \mathcal{E}}.
$
Thus, it follows from Condition \ref{cond1} and $\lambda = o(\tau/\sqrt{s})$ that
\begin{align}\label{e038}
E[\|\bdelta_{\alpha_0^c}\|_11_{\mathcal{E}_0^c\cap \mathcal{E}}] \leq C\tau^{-1}s\lambda^2 P(\mathcal{E}_0^c\cap \mathcal{E})= o(s\lambda n^{-c_1}) .
\end{align}
Since $\lambda = o(\tau/\sqrt{s})$, the first term on the right hand side of (\ref{e037}) is negative for sufficiently large $n$. So it follows from (\ref{e037}) that conditioning on $\mathcal{E}_0^c\cap \mathcal{E}$,
$
\big|\|\bdelta_{\alpha_0}\|_1 - Cs\lambda \big|\leq Cs\lambda.
$
Hence, we obtain that
\[
E[\|\bdelta_{\alpha_0}\|_11_{\mathcal{E}_0^c\cap \mathcal{E}} ] \leq Cs\lambda E[1_{\mathcal{E}_0^c\cap \mathcal{E}}] = O(s\lambda  n^{-c_1}).
\]
This together with (\ref{e038}) proves (\ref{e039}), which completes the proof of (\ref{e063}). Consequently, (\ref{e028}) follows and the result under the $L_2$-loss is proved.

We now consider  $E \|\bdelta\|_1$ under the $L_1$-estimation loss by using the following decomposition
\begin{align}\label{e070}
E \|\bdelta\|_1 = E[\|\bdelta\|_11_{\mathcal{E}_1}]+E[\|\bdelta\|_11_{\mathcal{E}\cap \mathcal{E}_0^c}]+E[\|\bdelta\|_11_{ \mathcal{E}^c}].
\end{align}
First, by Theorem \ref{Thm2}, the first term on the right hand side of (\ref{e070}) can be bounded as
\begin{align}\label{e027}
E[\|\bdelta\|_11_{\mathcal{E}_1}]\leq s\sqrt{(\log n)/n} P(\mathcal{E}_1) \leq Cs\sqrt{(\log n)/n}.
\end{align}
The second term of (\ref{e070}) has already been considered in (\ref{e039}). So we only need to study the third term. Since $\|\bdelta\|_1 \leq \sqrt{\kappa_c}\|\bdelta\|_2$, by (\ref{e040}) and (\ref{e035}), we can bound the third term as
\begin{align}\label{e072}
 E[\|\bdelta\|_11_{\mathcal{E}^c}] \leq\sqrt{\kappa_c}E[\|\bdelta\|_21_{\mathcal{E}^c}] \leq C\kappa_c E[\|n^{-1}\bX\t\bveps\|_\infty1_{\mathcal{E}^c}] + C\kappa_c \lambda P(\mathcal{E}^c)
=  O(p^{-c_1/2}\kappa_c).
\end{align}
Since $c_1$ can be chosen arbitrarily large and $\kappa_c \leq n + 1$, the above inequality together with (\ref{e027}), (\ref{e039}), and (\ref{e070}) leads to
\begin{align}\label{e055}
E \|\bdelta\|_1 \leq Cs(\log n)/n.
\end{align}
Thus, the risk result under the $L_1$-estimation loss is proved.

Now, applying H\"{o}lder's inequality and by (\ref{e028}) and (\ref{e055}), we can prove the risk inequalities under the $L_q$-estimation loss with $q \in (1,2)$, as in (\ref{e029}).

Finally we consider the $L_\infty$-estimation loss.
By (\ref{e028}) and Condition \ref{cond1},
\begin{align*}
E[\|\bdelta\|_{\infty}1_{\mathcal{E}_1^c}]\leq E[\|\bdelta\|_21_{\mathcal{E}_1^c}] \leq \{E[\|\bdelta\|_2^2]P\big(\mathcal{E}_1^c\big)\}^{1/2} = O\left\{s^{1/2}n^{-(c_1+1)/2}\sqrt{\log n}\right\}.
\end{align*}
Moreover, by Theorem \ref{Thm2}, we have $\|\bdelta\|_\infty 1_{\mathcal{E}_1}\leq C\gamma_n^*\sqrt{(\log n)/n}$. Since  $c_1$ can be chosen arbitrarily large, it follows that
\begin{align*}
E \|\bdelta\|_{\infty} =E[\|\bdelta\|_{\infty}1_{\mathcal{E}_1}] +E[\|\bdelta\|_{\infty}1_{\mathcal{E}_1^c}] \leq C\gamma_n^*\sqrt{(\log n)/n},
\end{align*}
which completes the proof for estimation risks.

\medskip

{\noindent \bf Prediction risk:} By (\ref{e031}) and Condition \ref{cond2}, we have
\begin{align}\label{e056}
 E \left\{D(\hbbeta)\right\}  = E[2^{-1}\bdelta\t\bX\t\bH(\tbbeta)\bX\bdelta] \leq (2c_2)^{-1}(I_1+I_2 + I_3),
\end{align}
where $I_1  = E[\|\bX\bdelta\|_2^21_{\mathcal{E}_1}]$, $I_2 = E[\|\bX\bdelta\|_2^21_{\mathcal{E}_0^c\cap\mathcal{E}}]$, and  $I_3 =E[\|\bX\bdelta\|_2^21_{\mathcal{E}^c}]$. We first consider $I_1 = E[\|\bX\bdelta\|_2^21_{\mathcal{E}_1}]$. By the second inequality in (\ref{e071}) and (\ref{e055}),
\begin{align}\label{e043}
I_1 \leq C\sqrt{(\log n)n}E[\|\bdelta\|_1 1_{\mathcal{E}_1}]\leq C\sqrt{(\log n)n}E[\|\bdelta\|_1]\leq Cs(\log n).
\end{align}
Next, we study the term $I_2$. By Lemma \ref{Lem1}, the definition of $\mathcal{E}$, and (\ref{e039}), we have
\begin{align}\label{e074}
I_2 \leq  Cn\lambda E[\|\bdelta\|_11_{\mathcal{E}_0^c\cap\mathcal{E}}] = O(s\lambda^2 n^{1-c_1}).
\end{align}
Now we consider the last term $I_3$. It follows from the proof of Lemma \ref{Lem1} that
\begin{align}\label{e075}
I_3 \leq  CE[|\bveps\t\bX\bdelta|1_{\mathcal{E}^c}] + Cn\lambda E[\|\bdelta\|_11_{\mathcal{E}^c}] \equiv I_{3,1}+I_{3,2}.
\end{align}
Since $\|\bdelta\|_1 \leq \sqrt{\kappa_c}\|\bdelta\|_2$, by (\ref{e062}) and (\ref{e032}), we can bound $I_{3,1}$ as
\begin{align}
\nonumber
I_{3,1}&=E[|\bveps\t\bX\bdelta|1_{\mathcal{E}^c}] \leq C E[\|\bdelta\|_1\|\bX\bveps\|_{\infty}1_{\mathcal{E}^c}]\leq C\{E[\|\bdelta\|_1^21_{\mathcal{E}^c}]\}^{1/2} \{E[\|\bX\bveps\|_{\infty}^21_{\mathcal{E}^c}]\}^{1/2}\\
& \leq C\sqrt{\kappa_c}\{E[\|\bdelta\|_2^21_{\mathcal{E}^c}]\}^{1/2}\{E[\|\bX\t\bveps\|_{\infty}^2 1_{\mathcal{E}^c}]\}^{1/2} = O(np^{-c_1/2}\kappa_c). \label{e005}
\end{align}
By (\ref{e072}), we have $I_{3,2} = O(\lambda np^{-c_1/2}\kappa_c)$. This together with (\ref{e075}) and (\ref{e005}) entails
\begin{align}\label{e018}
I_3 = O(np^{-c_1/2}\kappa_c).
\end{align}
Combing (\ref{e018}) with (\ref{e056})--(\ref{e074}) and noting that $c_1$ can be chosen arbitrarily large, we finish the proof for prediction risk.

\medskip

{\noindent \bf Sign risk:}  Since $\hbbeta \in \mathcal{B}_{\tau}$ and $p_{\lambda }(t)$ is increasing in $t\in [0, \infty)$, we have
$\|p_{\lambda }(\hbbeta)\|_1=\sum_{j=1}^pp_{\lambda }(|\hbeta_j|) \geq \|\hbbeta\|_0p_{\lambda }(\tau)$ and thus $\|\hbbeta\|_0 \leq [p_{\lambda }(\tau)]^{-1}\|p_{\lambda }(\hbbeta)\|_1$. This together with (\ref{e001}) and Condition \ref{cond2} gives
\begin{align}\label{e077}
\nonumber \FS(\hbbeta)&\leq \|\hbbeta\|_0+s \leq s+ [p_{\lambda }(\tau)]^{-1}\big[\|p_{\lambda }(\bbeta_0)\|_1 + n^{-1}\bveps\t\bX\bdelta - \frac{1}{2n}\bdelta\t\bX\t\bH(\tbbeta)\bX\bdelta\big]\\
&\leq s+[p_{\lambda }(\tau)]^{-1}\big[\|p_{\lambda }(\bbeta_0)\|_1 + n^{-1}\bveps\t\bX\bdelta \big].
\end{align}
Since $|n^{-1}\bveps\t\bX\bdelta| \leq \|n^{-1}\bX\t\bveps\|_\infty\|\bdelta\|_1 \leq \lambda \|\bdelta\|_1$ on the event $\mathcal{E}$, by (\ref{e005}) and (\ref{e039}) we have
\begin{align}\label{e078}
\nonumber & E[n^{-1}|\bveps\t\bX\bdelta|1_{\mathcal{E}_1^c}]= E[n^{-1}|\bveps\t\bX\bdelta|1_{\mathcal{E}^c}] + E[n^{-1}|\bveps\t\bX\bdelta|1_{\mathcal{E}_0^c\cap \mathcal{E}}] \\
&\leq O(p^{-c_1/2}\kappa_c) + \lambda E[\|\bdelta\|_11_{\mathcal{E}_0^c\cap \mathcal{E}}] = O(p^{-c_1/2}\kappa_c) + O(s\lambda^2n^{-c_1}).
\end{align}
 Thus, combining (\ref{e077}) with (\ref{e078}) and noting $\|p_{\lambda}(\bbeta_0)\|_1 \geq s p_{\lambda}(\tau)$, we obtain
\begin{align*}
E[\FS(\hbbeta)1_{\mathcal{E}_1^c}]& \leq P(\mathcal{E}_1^c)\left\{s+ [p_{\lambda }(\tau)]^{-1}\|p_{\lambda }(\bbeta_0)\|_1\right\} + [p_{\lambda }(\tau)]^{-1}E[n^{-1}|\bveps\t\bX\bdelta|1_{\mathcal{E}_1^c}] \\
&=  [p_{\lambda }(\tau)]^{-1}\big[ \|p_{\lambda }(\bbeta_0)\|_1 O(n^{-c_1}) +O(p^{-c_1/2}\kappa_c)+ O(s\lambda^2n^{-c_1})\big].
\end{align*}
On the other hand, Theorem \ref{Thm2} shows that $\FS(\hbbeta)=0$ on the event $\mathcal{E}_1$. Thus, we have $E[\FS(\hbbeta)1_{\mathcal{E}_1}]=0$, which leads to $E[\FS(\hbbeta)]= E[\FS(\hbbeta)1_{\mathcal{E}_1^c}]$. This concludes the proof.

\subsection{Proof of Theorem \ref{Thm4}}  \label{SecB.7}
To simplify the technical presentation, we first consider the case of linear model. Then the penalized negative log-likelihood minimization problem in (\ref{e177}) becomes the penalized least-squares problem with $Q_n(\bbeta) = (2n)^{-1}\|\by - \bX\bbeta\|_2^2 + \|p_\lambda(\bbeta)\|_1$. Note that in the case of linear model, $\bmu(\btheta) = \btheta$ and thus
\[ \left\|n^{-1} \bX_\alpha\t \left[\bmu(\bX \bbeta) - \bmu(\bX \bbeta_0)\right]\right\|_2 = \|n^{-1} \bX_\alpha\t \bX (\bbeta - \bbeta_0)\|_2 \geq c_4 \|\bbeta - \bbeta_0\|_2 \]
holds for any $\bbeta \in \mathcal{B}_{\tau}$, with $c_4 = c^2$ and $\alpha = \supp(\bbeta) \cup \supp(\bbeta_0)$. Denote by $\bdelta = (\delta_1, \cdots, \delta_p)\t= \hbbeta - \bbeta_0$ with $\hbbeta = (\hbeta_1, \cdots, \hbeta_p)\t$. Let $\alpha_0 = \supp(\bbeta_0)$ and $\alpha_1=\supp(\hbbeta)$. Clearly, $\supp(\bdelta) \subset \alpha = \alpha_0 \cup \alpha_1$. It follows from $\bbeta_0, \hbbeta \in \mathcal{B}_{\tau}$ that $|\alpha_0|<\kappa_c/2$, $|\alpha_1| < \kappa_c/2$, and $|\alpha| \leq |\alpha_0| + |\alpha_1| < \kappa_c$. Thus by the definition of the robust spark $\kappa_c$, we have $\lambda_{\min}(n^{-1}\bX_\alpha\t\bX_\alpha) \geq c^2$, which leads to
\begin{equation}\label{e080}
\|\bdelta\|_2 = \|\bdelta_\alpha\|_2 \leq c^{-2}\|n^{-1}\bX_\alpha\t \bX_\alpha\bdelta_\alpha\|_2 = c^{-2}\|n^{-1}\bX_\alpha\t \bX \bdelta\|_2.
\end{equation}
Since $\by = \bX \bbeta_0 + \bveps$ in linear model, we have $\bX \bdelta = \bX(\hbbeta - \bbeta_0) = - (\by - \bX\hbbeta) + \bveps$ and thus
\begin{align*}
n^{-1}\bX_\alpha\t\bX\bdelta = n^{-1}\bX_\alpha\t \left[-(\by - \bX\hbbeta) + \bveps\right] = -n^{-1}\bX_\alpha\t(\by - \bX\hbbeta) + n^{-1}\bX_\alpha\t\bveps.
\end{align*}
This representation together with (\ref{e080}) yields
\begin{equation}\label{e083}
\|\bdelta\|_2 \leq c^{-2}\|n^{-1}\bX_\alpha\t \bX \bdelta\|_2 \leq c^{-2}\|n^{-1}\bX_\alpha\t(\by - \bX\hbbeta)\|_2 + c^{-2}\|n^{-1}\bX_\alpha\t\bveps\|_2.
\end{equation}
Such an inequality provides an effective way to bound the size of the set $\alpha$.

By Condition \ref{cond1}, the event $\mathcal{E} = \{\|n^{-1} \bX\t \bveps\|_\infty \leq \lambda/2\}$ has a large probability. We condition on this event hereafter. Then it holds that
\begin{equation} \label{e084}
\|n^{-1}\bX_\alpha\t\bveps\|_2 \leq |\alpha|^{1/2} \|n^{-1}\bX_\alpha\t\bveps\|_\infty \leq |\alpha|^{1/2} \lambda/2.
\end{equation}
Since $\eta_n = \|n^{-1}\bX\t[\by - \bmu(\bX\hbbeta)]\|_\infty = \|n^{-1}\bX\t(\by - \bX\hbbeta)\|_\infty$, we have
\begin{equation} \label{e085}
\|n^{-1}\bX_\alpha\t(\by - \bX\hbbeta)\|_2 \leq |\alpha|^{1/2} |n^{-1}\bX_\alpha\t(\by - \bX\hbbeta)\|_\infty \leq |\alpha|^{1/2} \eta_n.
\end{equation}
Let $k = |\alpha_1 \setminus \alpha_0|$. Clearly, $|\alpha| = |\alpha_0| +  |\alpha_1 \setminus \alpha_0| = s + k$. Note that for each $j \in \alpha_1 \setminus \alpha_0$, we have $\delta_j = \hbeta_j-\beta_{0,j} = \hbeta_j$ and thus $|\delta_j| \geq \tau$, which entails that
\begin{equation}\label{e081}
\|\bdelta\|_2 \geq k^{1/2} \tau.
\end{equation}
Combining inequalities (\ref{e083})--(\ref{e081}) along with $|\alpha| = s + k$ gives
\[
k^{1/2}\tau \leq c^{-2}(s+k)^{1/2}(\eta_n + 2^{-1} \lambda),
\]
which ensures that
\begin{equation} \label{e086}
k\leq \frac{c^{-4}(\eta_n + 2^{-2} \lambda)^2/\tau^2}{1-c^{-4}(\eta_n + 2^{-2} \lambda)^2/\tau^2} s.
\end{equation}
Since $\eta_n + \lambda=o(\tau)$, it follows from the bound in (\ref{e086}) that $k \leq s$ for large enough $n$. Thus, applying similar arguments as above results in
\begin{equation} \label{e087}
\|\bdelta\|_2 \leq c^{-2} (2 s)^{1/2} (\eta_n + 2^{-1} \lambda).
\end{equation}
Since $\min_{1\leq j\leq s}|\beta_{0,j}| > c_5 s^{1/2} (\eta_n+\lambda)$ with $c_5$ some sufficiently large positive constant, the above inequality (\ref{e087}) entails that for large enough $n$, $\hbeta_j \neq 0$ for each $j \in \alpha_0$. This shows that $\supp(\hbbeta) \supset \alpha_0 = \supp(\bbeta_0)$. Note that by assumption, $\hbbeta$ is the global minimizer of the problem (\ref{e177}) when constrained on the subspace given by its support. Observe that all arguments in the proofs of Theorems \ref{Thm1}--\ref{Thm3} on the global minimizer equally apply to the computable solution $\hbbeta$ as long as $\supp(\hbbeta) \supset \supp(\bbeta_0)$. Therefore, $\hbbeta$ enjoys the same asymptotic properties as for any global minimizer in Theorems \ref{Thm1}--\ref{Thm3} under the same conditions therein.

For the case of nonlinear model, by assumption we have
\[ \left\|n^{-1} \bX_\alpha\t \left[\bmu(\bX \hbbeta) - \bmu(\bX \bbeta_0)\right]\right\|_2 \geq c_4 \|\bdelta\|_2, \]
which together with $\bveps = \by - \bmu(\bX \bbeta_0)$ leads to
\begin{equation} \label{e088}
\|\bdelta\|_2 \leq c_4^{-1} \left\|n^{-1}\bX_\alpha\t \left[\by - \bmu(\bX\hbbeta)\right]\right\|_2 + c_4^{-1}\|n^{-1}\bX_\alpha\t\bveps\|_2.
\end{equation}
Observe that inequality (\ref{e088}) is of similar form as (\ref{e083}). Thus an application of similar arguments as above completes the proof.

\end{document}